\documentclass{taira}

\usepackage{graphicx}
\usepackage{newtxtext}
\usepackage{newtxmath}
\usepackage{natbib}
\usepackage{hyperref}
\hypersetup{
	colorlinks = true,
	urlcolor   = blue,
	citecolor  = black,
}

\newcommand{\RomanNumeralCaps}[1]
\linenumbers

\usepackage{graphicx,amssymb,amsmath,color,epsfig,soul,rotating,overpic,varwidth,xcolor}

\usepackage[export]{adjustbox}
\usepackage{subfigure}
\usepackage{subfigmat}
\usepackage{algorithm}
\usepackage[noend]{algpseudocode}
\usepackage{bm}
\usepackage{xfrac}
\usepackage{multicol}
\usepackage{multirow}
\usepackage{pbox}
\usepackage{tabularx}

\usepackage{caption}

\usepackage{transparent}

\usepackage{booktabs}
\usepackage{tikz}
\usetikzlibrary{tikzmark}

\usepackage{lipsum}

\definecolor{blue2}{rgb}{0, 0.4470, 0.7410}
\definecolor{red2}{rgb}{0.8500, 0.1250, 0.0480} 
\definecolor{orange2}{rgb}{0.8500, 0.3250, 0.0980} 
\definecolor{yellow2}{rgb}{0.9290, 0.6940, 0.1250}
\definecolor{purple2}{rgb}{0.4940, 0.1840, 0.5560}
\definecolor{green2}{rgb}{0.4660, 0.6740, 0.1880}
\definecolor{ltblue2}{rgb}{0.3010, 0.7450, 0.9330}
\definecolor{dkred2}{rgb}{0.6350, 0.0780, 0.1840}
\definecolor{gray2}{rgb}{0.22, 0.22, 0.3}
\definecolor{blueIV}{rgb}{0, 0, 0.7410}
\definecolor{blueIII}{rgb}{0.2, 0.2, 0.7410}
\definecolor{blueII}{rgb}{0.4, 0.4, 0.7410}
\definecolor{blueI}{rgb}{0.7410, 0.7410, 0.7410}
\definecolor{jetVI}{rgb}{0.9763    0.9831    0.0538}
\definecolor{jetV}{rgb}{0.9264    0.7256    0.2996}
\definecolor{jetIV}{rgb}{0.4783    0.7489    0.4877}
\definecolor{jetIII}{rgb}{0.0282    0.6663    0.7574}
\definecolor{jetII}{rgb}{0.0582    0.4677    0.8589}
\definecolor{jetI}{rgb}{0.2081    0.1663    0.5292}

\newcommand{\bs}{\boldsymbol}

\begin{document}
	
	\captionsetup{font=scriptsize,labelfont=scriptsize}
	
	\shorttitle{Wing sweep effects on laminar separated flows} 
	\shortauthor{J. H. M. Ribeiro et al.} 
	
	\title{Wing sweep effects on laminar separated flows} 
	
	\author
	{
		Jean H\'{e}lder Marques Ribeiro\aff{1}
		\corresp{\email{jeanmarques@g.ucla.edu}},
		Chi-An Yeh\aff{1,2},
		Kai Zhang\aff{3}
		\corresp{Current affiliation: School of Naval Architecture, Ocean and Civil Engineering, Shanghai Jiao Tong University, Shanghai 200240, China},
		\and
		Kunihiko Taira\aff{1}
	}
	
	\affiliation
	{
		\aff{1}
		Department of Mechanical and Aerospace Engineering, University of California, Los Angeles, CA 90095, USA
		
		\aff{2}
		Department of Mechanical and Aerospace Engineering, North Carolina State University, Raleigh, NC 27695, USA
		
		\aff{3}
		Department of Mechanical and Aerospace Engineering, Rutgers University, Piscataway, NJ 08854, USA
		
	}
	
	\maketitle
	
	\begin{abstract}
		
		We reveal the effects of sweep on the wake dynamics around NACA 0015 wings at high angles of attack using direct numerical simulations and resolvent analysis. The influence of sweep on the wake dynamics is considered for sweep angles from $0^\circ$ to $45^\circ$ and angles of attack from $16^\circ$ to $30^\circ$ for a spanwise periodic wing at a chord-based Reynolds number of $400$ and a Mach number of $0.1$. Wing sweep affects the wake dynamics, especially in terms of stability and spanwise fluctuations with implications on the development of three-dimensional wakes. We observe that wing sweep attenuates spanwise fluctuations. Even as the sweep angle influences the wake, force and pressure coefficients can be collapsed for low angles of attack when examined in wall-normal and wingspan-normal independent flow components. Some small deviations at high sweep and incidence angles are attributed to vortical wake structures that impose secondary aerodynamic loads, revealed through the force element analysis. Furthermore, we conduct global resolvent analysis to uncover oblique modes with high disturbance amplification. The resolvent analysis also reveals the presence of wavemakers in the shear-dominated region associated with the emergence of three-dimensional wakes at high angles of attack.  For flows at high sweep angles, the optimal convection speed of the response modes is shown to be faster than the optimal wavemakers speed suggesting a mechanism for the attenuation of perturbations. The present findings serve as a fundamental stepping stone to understanding separated flows at higher Reynolds numbers.
		
	\end{abstract}

	\section{Introduction}
	\label{sec:intro}
	
	Understanding the dynamics of airfoil wakes is critically important for the design of aircraft. Moreover, many nature-inspired engineering applications can benefit from the study of the complex fluid dynamics observed, for instance, in the flight of common swifts (\textit{Apus Apus}), where wings are swept \citep{Videler:Science04}. The dynamics of wakes have been studied extensively to reveal the mechanisms that trigger flow separation and three-dimensionality over unswept wings \citep{Anderson:10,Taira:JFM09,Zhang:JFM20}. The wake dynamics of swept wings, however, have not received much attention to understand the effect of sweep on the vortical structures that emerge at high angles of attack. 
	
	Fundamental studies on flow separation have been performed on two-dimensional (2D) unswept wings. The flow structures emerging in post-stall wakes have been a subject of research for decades \citep{Gaster:67,Tobak:AR82}. In the early work of \cite{Horton:68}, the behavior of a canonical laminar boundary layer separation was investigated through theoretical and experimental approaches. On the numerical side, $2$D simulations of flow separation were performed by \cite{Pauley:JFM90}, establishing a relation among vortex shedding, adverse pressure gradient, and inviscid shear layer mechanisms.
	
	The characteristics of vortex shedding are related to geometrical parameters of the wing and physical parameters of the flow, including the angle of attack and the Reynolds number \citep{Huang:JFM01,Yarusevych:JFM09}. The Reynolds number is important for discussing the transition on vortex shedding patterns in $2$D laminar flows \citep{Williamson:JFS88,Rossi:JFM18}. For the analysis of flow fields at the Reynolds number where such transitions occur, experiments and computations have shown that three-dimensionality emerges as stall cells develop on the suction side  \citep{Winkelman:AIAAJ80}. Numerically, three-dimensionality at high angles of attack can be captured by extending the wingspan in spanwise periodic simulations \citep{Braza:JFM01,Hoarau:JFM03}.
	
	Around swept wings at high incidence, vortical structures are affected by the combination of the streamwise and spanwise flows, where the latter yields a crossflow instability over the airfoil \citep{Serpieri:JFM16}. The spanwise flow alters the stall characteristics and vortical interactions in the airfoil wake \citep{Harper:64}. Laminar flows over swept wings have been examined through experiments \citep{Yen:AIAAJ07} and numerical simulations \citep{Mittal:JFS14,Mittal:JFM21,Zhang:JFM20b,Zhang:PRF22}. Turbulent flows over swept wings have also been studied through large eddy simulations \citep{Visbal:AIAAJ19,Garmann:AIAA20}. Such efforts, however, have considered finite swept wings, hence the effects of sweep angle are not independently analyzed from the wing tip effects. 
	
	To distinguish the influence of sweep from tip effects, one may consider analyzing a spanwise periodic swept wing, as in the works of \cite{Paladini:PRF19}, \cite{Crouch:JFM19}, and \cite{Plante:AIAAJ20,Plante:JFM21}, which revealed stall cells advection during transonic buffet over swept wings. Although the wake dynamics is influenced by the sweep angle, the flows of swept and unswept wings still preserve similarities in the chordwise and wall normal flow components through the boundary layer independence principle \citep{White91,Wygnanski:JFM11}. This principle has prompted many studies in turbulent flow regimes \citep{Wygnanski:JA14,Coleman:JFM19}, although the independence principle for laminar separated flows remains to be examined.
	
	Analyzing the flow variables on the plane normal to the leading edge, we are able to identify the collapse of laminar flow characteristics for flows over swept wings at lower angles of attack, suggesting an independence  of streamwise and spanwise flow components. When interaction between them is present, it is not expected that the independence principle holds. In this study, we call on the force element theory \citep{Chang:PRSA92} to reveal the flow structures that exert additional forces on the wing responsible for the departure from the independence principle.
	
	The presence of spanwise instabilities in the wakes behind swept wings suggests the existence of self-sustained mechanisms that affect the wake dynamics. For instance, these mechanisms may be responsible for initiating three-dimensionality at higher incidence and reducing spanwise oscillations in swept wings. This flow complexity motivates the use of modal analysis \citep{Taira:AIAAJ17,Taira:AIAAJ20} to provide a comprehensive understanding of the wake dynamics and evolution of disturbances in swept wings.
	
	Among all modal analysis methods, the resolvent analysis reveals the evolution of perturbations excited by optimal harmonic inputs to the flow \citep{Trefethen:93,Farrell:PRL94,Jovanovic:JFM05}. Resolvent analysis can be performed with respect to steady (equilibrium) and time-averaged states. The latter case assumes that the flow is statistically stationary. In such a case, resolvent analysis can be used to study laminar and turbulent flows, extending the applicability of resolvent analysis to time-averaged base flows \citep{McKeon:JFM10}. \cite{Jovanovic:04} extended the methodology to unstable systems and \cite{Schmid:AMR14} discussed the evolution of perturbations over a finite-time horizon and the modal sensitivity to small changes in the resolvent operator. These efforts enabled the use of resolvent analysis for studying various types of complex fluid flows \citep{Gomez:JFM16,Schmidt:JFM18,Yeh:JFM19,Yeh:PRF20,Kojima:JFM20,Liu:JFM21}. 
	
	For laminar separated flows, resolvent analysis reveals how flow perturbations arise, grow and self-sustain in the flow field. For instance, disturbances generated from the vortices at the flow separation over the wing can grow and develop into wake unsteadiness downstream. This behavior can be captured by the optimal forcing and response structures, and their spatial overlap, characterized by wavemakers. Such regions of the flow field act as a source to the global stability of the flow and are optimal locations for the introduction of self-sustained perturbations in the flow field \citep{Giannetti:JFM07,Giannetti:JFM10,FosasdePando:JFM17}. 
	
	Wavemaker analysis can aid in uncovering mechanisms that sustain flow unsteadiness in particular flow regions. In the present work, we further reveal that optimal response structures have a lower phase speed than the optimal wavemakers, which explains the attenuation of unsteadiness and three-dimensionality in flows over swept wings. Furthermore, resolvent analysis predicts the onset of oblique shedding on flows over swept wings, as observed in flows over high-apect-ratio bodies \citep{Mittal:JFS14,Mittal:JFM21,Zhang:JFM20b}.  As oblique vortices are observed in laminar flows over finite-length bodies and unseen over spanwise periodic bodies, there is an open question on whether sweep angle or the body tip promotes oblique shedding. In the present work, we address this question using revolvent analysis. The emergence of highly amplified oblique resolvent modes shows that oblique shedding can be triggered and sustained over swept wings with an appropriate input.
	
	This work aims to study the influence of the sweep angle on the wake dynamics of laminar flows over swept wings by combining direct numerical simulations and resolvent analysis. We present the problem setup and the numerical methods in section \ref{sec:problem}.  Next, we characterize the flow over swept wings in section \ref{sec:swept}. We also examine the applicability of the  concepts associated with the boundary layer independence principle for flows with massive separation and employ the force element theory to identify sources of vortically induced lift and drag in sections \ref{sec:wake} and \ref{sec:forceelements}. Moreover, we discuss the effects of spanwise flow on the evolution of perturbations via resolvent analysis in section \ref{sec:resolvent}. The role of wavemakers in swept wings is also studied in sections \ref{sec:wavemakers}. Lastly, we summarize our findings \ref{sec:conclusions}.
	
	\section{Problem setup}
	\label{sec:problem}
	
\begin{figure}
		\footnotesize
		\centering
		\begin{tikzpicture}
		\node[anchor=south west,inner sep=0] (image) at (-0.2,0.0) {\includegraphics[trim=0mm 2mm 2mm 0mm, clip,width=1\textwidth]{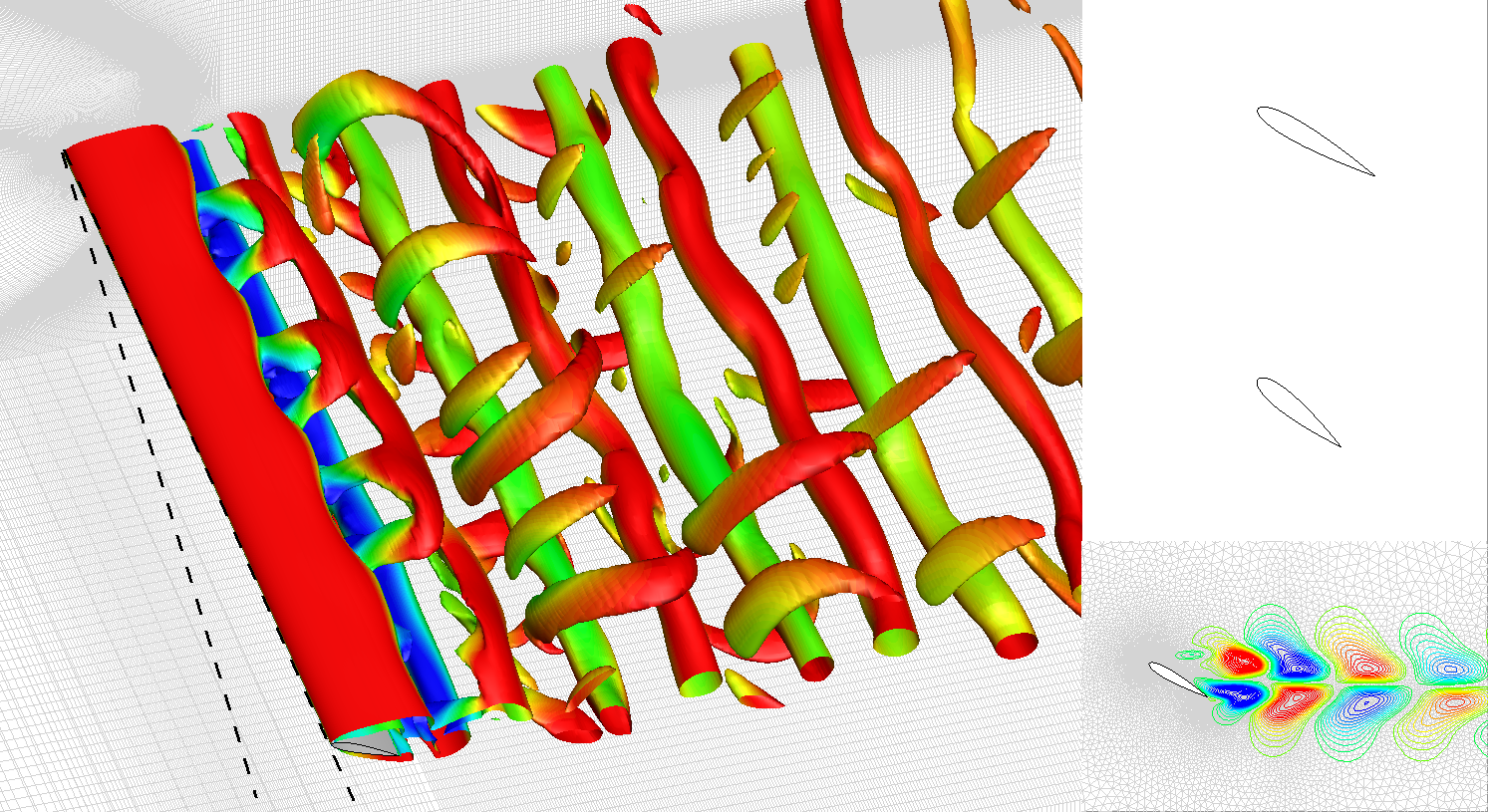}};
		\draw[->,black, very thick] (0.35,1.4) -- (0.35,2.45);     
		\draw[->,black, very thick] (0.35,1.4) -- (0.65,0.4);     
		\draw[->,black, very thick] (0.35,1.4) -- (1.35,1.60);    
		\draw[->,black, thick,dashed] (0.35,1.4) -- (1.20,1.85); 
		\draw[->,black, thick,dashed] (0.35,1.45) -- (0.85,0.5);   
		\node[text width=0cm] at (0.05,2.5) {$y$};
		\node[text width=0cm] at (0.2,0.4) {$z$};
		\node[text width=0cm] at (1.00,0.7) {$z'$};
		\node[text width=0cm] at (1.30,1.3) {$x$};
		\node[text width=0cm] at (1.15,2.1) {$x'$};
		\draw[<->,black, thick] (2.0,0.5) arc (280:290:4.5);
		\node[text width=0cm] at (2.4,0.3) {$\Lambda$};
		\scriptsize
		\draw[->,black, thick] (12.5,6.5) -- (12.5,7.0);
		\draw[->,black, thick] (12.5,6.5) -- (13.0,6.5);
		\draw[<->,black, thick] (11.1,6.05) -- (12.2,5.45);
		\draw[->,black, thick] (10.1,6.35) -- (11.27,6.35);
		\draw[black, thick,dashed] (10.3,6.9) -- (11.27,6.35);
		\draw[<->,black, thick] (10.3,6.35) arc (180:150:1.0);
		\node[text width=0cm] at (12.2,7.0) {$y$};
		\node[text width=0cm] at (13.0,6.3) {$x$};
		\node[text width=0cm] at (10.0,6.65) {$\alpha$};
		\node[text width=0cm] at (10.4,6.1) {$U_\infty$};
		\node[text width=0cm] at (11.2,5.6) {$L_c$};
		\scriptsize
		\draw[->,black, thick] (12.5,4.0) -- (12.5,4.5);
		\draw[->,black, thick] (12.5,4.0) -- (13.0,4.0);
		\draw[<->,black, thick] (11.1,3.65) -- (11.95,3.05);
		\draw[->,black, thick] (10.1,3.85) -- (11.27,3.85);
		\draw[black, thick,dashed] (10.3,4.4) -- (11.27,3.85);
		\draw[<->,black, thick] (10.3,3.85) arc (180:150:1.0);
		\node[text width=0cm] at (12.2,4.5) {$y$};
		\node[text width=0cm] at (13.0,3.8) {$x'$};
		\node[text width=0cm] at (9.85,4.2) {$\alpha_\text{eq}$};
		\node[text width=0cm] at (9.9,3.65) {$U_\infty \cos \Lambda$};
		\node[text width=0cm] at (11.1,3.1) {$L_c^\prime$};
		\draw[rounded corners,gray, ultra thick] (-0.2,0.0) rectangle (9.65,7.38);
		\draw[rounded corners,gray, fill=white, ultra thick] (0.5,7.18) rectangle (2.9,7.58);
		\node[align=left] at (1.7,7.38) {DNS domain};
		\draw[rounded corners,gray, ultra thick] (9.7,2.39) rectangle (13.3,7.38);
		\draw[rounded corners, gray, fill=white, ultra thick] (9.9,7.18) rectangle (12.3,7.58);
		\node[align=left] at (11.1,7.38) {$2$D slice};
		\draw[rounded corners,blue2, ultra thick] (9.7,0.0) rectangle (13.3,2.38);
		\draw[rounded corners, blue2, fill=white, ultra thick] (9.9,2.18) rectangle (12.3,2.58);
		\node[align=left] at (11.1,2.38) {resolvent domain};
		\end{tikzpicture} \\ \vspace{-2mm}
		\caption{The problem setup. Instantaneous flow over a spanwise periodic NACA 0015 profile at $\alpha = 30^\circ$ and sweep angle $\Lambda = 15^\circ$ visualized with isosurfaces of $Q$ criterion colored by streamwise velocity $u_x$. $2$D slice of airfoil in coordinate systems ($x,y,z$) and ($x',y,z'$), with $x'$  and $z'$ perpendicular  and parallel to the wingspan, respectively. DNS and resolvent grids are shown as gray lines in the background.} 
		\label{fig:phys_setup}
	\end{figure}
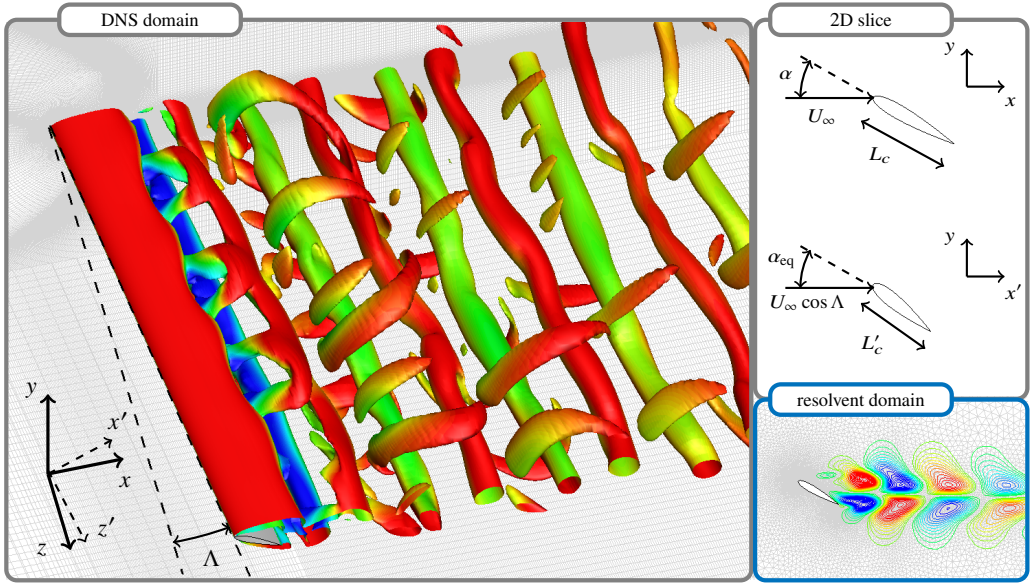

	We study laminar separated flows over swept wings with a NACA 0015 airfoil cross-section for sweep angles $0^\circ \leq \Lambda \leq 45^\circ$ and angles of attack $16^\circ \leq \alpha \leq 30^\circ$. For all cases, we set the chord-based Reynolds number to $Re_{c} \equiv U_\infty L_c / \nu = 400$ and the free-stream Mach number to $M_\infty \equiv U_\infty / a_\infty = 0.1$. Here, $U_\infty$ is the free-stream velocity, $L_c$ is the chord-length, $\nu$ is the kinematic viscosity and $a_\infty$ is the free-stream speed of sound. We illustrate the present setup in figure \ref{fig:phys_setup} with an instantaneous flow field visualized for $\alpha = 30^\circ$ and $\Lambda = 15^\circ$. 
	
	For the present work, we consider a NACA 0015 profile with constant chord-length in the $(x,y)$ plane for all angles of attack and sweep, with spanwise periodicity in the $z^\prime$-direction. As shown in figure \ref{fig:phys_setup}, the effective chord-length is defined as $L_{c}^\prime = L_c  \left( \cos^2 \alpha \cos^2 \Lambda + \sin^2 \alpha \right)^{1/2} \le L_c$ and the effective angle of attack is defined as $\alpha_{\text{eq}} = \tan^{-1} \left( \tan\alpha / \cos\Lambda \right) \ge \alpha$. Effective $L_c^\prime$ and $\alpha_{\text{eq}}$ are dependent on the sweep angle $\Lambda$ and relate to the airfoil geometry on the $(x',y)$ plane. 
	
	\subsection{Direct numerical simulations}
	\label{sec:dns_computational}
	
	To study the flows over  swept NACA 0015 airfoils, we perform direct numerical simulations (DNS) with \textit{CharLES}, a finite-volume-based compressible flow solver with second- and third-order accuracies in space and time, respectively \citep{Khalighi:AIAA11,Bres:AIAAJ17}. We position the leading edge of the airfoil at $(x^\prime/L_c,\ y/L_c) = (0,0)$. The C-shaped computational mesh extends over $(x^\prime/L_c, y/L_c, z^\prime/L_c) \in [-20,25] \times  [-20,20] \times [0,4]$.  We build the grids with $(\min \Delta x, \min \Delta y, \min \Delta z)/L_c = (0.005, 0.005, 0.0625)$, with mesh refinement near the airfoil and wake. This yields a mesh with $100,000$ to $200,000$ cells on the root plane, extruded in the spanwise direction with equally spaced cell elements. We have verified our computational setup and validated the results with \cite{Zhang:JFM20b}. Our simulations obtained close agreement for the instantaneous and time-averaged velocity components, skin friction lines and pressure coefficients over the wing surface.
			
	We prescribe Dirichlet boundary conditions at the inlet and farfield boundaries as $(\rho, u_{x'}, u_y, u_{z'}, p) = (\rho_\infty, U_\infty \cos\Lambda, 0, U_\infty \sin\Lambda, p_\infty)$, where $\rho$ is density, $p$ is pressure, $u_{x^\prime}$, $u_y$, and $u_{z^\prime}$ are velocity components in ($x^\prime,y,z^\prime$) directions, respectively, $\rho_\infty$ is the freestream density and $p_\infty$ is the freestream pressure. On the airfoil surface, we prescribe the adiabatic no-slip boundary condition. For the outflow, a sponge layer \citep{Freund:AIAAJ97} is applied over $x^\prime/L_c \in [15,25]$ with the target state being the running-averaged flow over $5$ acoustic time units. For time integration, a fixed acoustic Courant-Friedrichs-Lewy (CFL) number of $1$ is utilized. We start the simulations with uniform flow. The initial transients are flushed out of the computational domain for $80$ convective time units, after which statistics are recorded over at least $100$ convective time units.
	
	\renewcommand{\arraystretch}{1.5}
	\begin{table}
		\centering
		\begin{tabular}{p{1.15in}p{0.37in}p{0.37in}p{0.001in}p{0.37in}p{0.37in}p{0.001in}p{0.37in}p{0.37in}p{0.001in}p{0.37in}p{0.37in}}
			\multicolumn{1}{r}{} &
			\multicolumn{2}{c}{\hspace{0mm}$\Lambda = 0^\circ$} &&
			\multicolumn{2}{c}{\hspace{0mm}$\Lambda = 15^\circ$} &&
			\multicolumn{2}{c}{\hspace{0mm}$\Lambda = 30^\circ$} &&
			\multicolumn{2}{c}{\hspace{0mm}$\Lambda = 45^\circ$} \\
			& 
			\multicolumn{1}{c}{\hspace{0mm}$\overline{C_L}$} & \multicolumn{1}{c}{\hspace{0mm}$\overline{C_D}$} && 
			\multicolumn{1}{c}{\hspace{0mm}$\overline{C_L}$} & \multicolumn{1}{c}{\hspace{0mm}$\overline{C_D}$} &&
			\multicolumn{1}{c}{\hspace{0mm}$\overline{C_L}$} & \multicolumn{1}{c}{\hspace{0mm}$\overline{C_D}$} &&
			\multicolumn{1}{c}{\hspace{0mm}$\overline{C_L}$} & \multicolumn{1}{c}{\hspace{0mm}$\overline{C_D}$} \\ 
			\cline{2-3}\cline{5-6}\cline{8-9}\cline{11-12}
			Present & $0.690$ & $0.405$ && $0.649$ & $0.391$ && $0.536$ & $0.350$ && $0.384$ & $0.296$ \\
			\cite{Zhang:JFM20b} & $0.702$ & $0.405$ && $0.657$ & $0.392$ && $0.547$ & $0.353$ && $0.393$ & $0.301$  \\
		\end{tabular}
		\caption{Time-averaged lift and drag coefficients ($\overline{C_L}$ and $\overline{C_D}$) compared to \cite{Zhang:JFM20b} for laminar separated flow over NACA 0015 airfoils with $\alpha = 20^\circ$, $\Lambda = 0^\circ$, $15^\circ$, $30^\circ$, and $45^\circ$. }
		\label{tab:validation}
	\end{table}
	
	The current results are carefully validated by examining the lift, drag, and pressure coefficients, respectively defined as
	\begin{equation}
	C_L = \frac{F_y}{\frac{1}{2}\rho U_\infty^2  L_c}, \quad C_D = \frac{F_x}{\frac{1}{2}\rho U_\infty^2  L_c}, \quad \text{and}  \quad C_p = \frac{p - p_\infty}{\frac{1}{2}\rho U_\infty^2 },
	\label{eq:cd_cl}
	\end{equation}
	where $F_x$ and $F_y$ are the $x$ and $y$ force components, respectively. The forces for $\alpha = 20^\circ$ are in close agreement with those reported by \cite{Zhang:JFM20b}, as shown in table \ref{tab:validation}. The flow fields were also compared and exhibit agreement validating the current setup.
	
	\subsection{Resolvent analysis}
	\label{sec:eigen_formulation}
	
	To analyze the perturbation dynamics over swept wings, let us decompose the flow state $\mathbf{q}$ as
	\begin{equation}
	\mathbf{q}= \bar{\mathbf{q}} + \mathbf{q}' \mbox{   ,}
	\label{eq:1}
	\end{equation}
	where $\bar{\mathbf{q}}$ is the time- and spanwise-averaged flow and $\mathbf{q}'$ is the fluctuating component. With this Reynolds decomposition, we can express the compressible Navier--Stokes equations as
	\begin{equation}
	\frac{\partial \mathbf{q}'}{\partial t} = \mathcal{L}_{\bar{\mathbf{q}}}\mathbf{q}' + \mathbf{f}' \mbox{   ,}
	\label{eq:LNS}
	\end{equation}
	where $\mathbf{f}'$ gathers all nonlinear terms. The spanwise periodicity and statistical stationarity allows for $\mathbf{q}'$ and $ \mathbf{f}'$ to be represented with temporal and spanwise Fourier modes
	\begin{equation}
	\left[ \mathbf{q}'(x',y,z',t), \mathbf{f}'(x',y,z',t) \right] = \int_{-\infty}^{\infty} \int_{-\infty}^{\infty} \left[ \hat{\mathbf{q}}_{k_{z^\prime},\omega}(x',y), \hat{\mathbf{f}}_{k_{z^\prime},\omega}(x',y) \right] e^{i \left(k_{z^\prime} z' - \omega t \right)} {\rm d} k_{z^\prime} {\rm d} \omega \mbox{   ,}
	\label{eq:fourier}
	\end{equation}
	where  $k_{z^\prime}$ is the spanwise wavenumber, and $\hat{\mathbf{q}}_{k_{z^\prime},\omega}$ and $\hat{\mathbf{f}}_{k_{z^\prime},\omega}$ are the biglobal modes for spanwise wavenumber $k_{z^\prime}$ and temporal frequency $\omega$. By substituting these modal expressions into equation \ref{eq:LNS}, we obtain
	\begin{equation}
	-i \omega \hat{\mathbf{q}}_{k_{z^\prime},\omega} = \mathcal{L}_{\bar{\mathbf{q}}}\hat{\mathbf{q}}_{k_{z^\prime},\omega} + \hat{\mathbf{f}}_{k_{z^\prime},\omega} \mbox{   ,}
	\label{eq:LNS_fourier}
	\end{equation}
	which is an inhomogeneous linear differential equation describing the perturbation evolution with input  $\hat{\mathbf{f}}_{k_{z^\prime},\omega}$. This formulation leads us to
	\begin{equation}
	\hat{\mathbf{q}}_{k_{z^\prime},\omega} = \mathcal{H}_{\bar{\mathbf{q}}(k_{z^\prime},\omega)} \hat{\mathbf{f}}_{k_{z^\prime},\omega} \mbox{   ,} 
	\label{eq:cont_resolvent}
	\end{equation}
	where
	\begin{equation}
	\mathcal{H}_{\bar{\mathbf{q}}(k_{z^\prime},\omega)} = \left[ -i \omega I  - \mathcal{L}_{\bar{\mathbf{q}}}(k_{z^\prime}) \right]^{-1}
	\label{eq:cont_resolvent2}
	\end{equation}
	is known as the resolvent \citep{Trefethen:93,Reddy:JFM93}. This operator acts as a state-space transfer function between the input forcing $\hat{\mathbf{f}}_{k_{z^\prime},\omega}$ and the output response $\hat{\mathbf{q}}_{k_{z^\prime},\omega}$ with respect to the base flow $\bar{\mathbf{q}}$ \citep{Jovanovic:JFM05}. To construct a discrete linear operator $\mathbf{L}_{\bar{\mathbf{q}}}$ and a discrete resolvent $\mathbf{H}_{\bar{\mathbf{q}}} \in \mathbb{C}^{m \times m}$, we use the time-averaged flow $\bar{\mathbf{q}}$ \citep{McKeon:JFM10}. Here, $m$ is the resolvent operator size defined by the product of the number of spatial grid points used to discretize the domain and the number of state variables. Appropriate boundary conditions are embedded in $\mathbf{H}_{\bar{\mathbf{q}}}$.
	
	The discrete resolvent is  examined through singular value decomposition (SVD),
	\begin{equation}
	\mathbf{H}_{\bar{\mathbf{q}}} = \left[ -i \omega \mathbf{I}  - \mathbf{L}_{\bar{\mathbf{q}}} \right]^{-1} = \mathbf{Q} \bs{\Sigma} \mathbf{F}^* ,
	\end{equation} 
	where $\mathbf{F} = [\hat{\mathbf{f}}_1,\hat{\mathbf{f}}_2,\dots,\hat{\mathbf{f}}_m]$ is the orthonormal matrix of right singular vectors representing the forcing modes, $\bs{\Sigma} = {\rm diag}[\sigma_1,\sigma_2,\dots,\sigma_m]$ is the diagonal matrix of singular values ranked in descending order, and $\mathbf{Q} =  [\hat{\mathbf{q}}_1,\hat{\mathbf{q}}_2,\dots,\hat{\mathbf{q}}_m]$ is the orthonormal matrix of left singular vectors representing the response modes \citep{Trefethen:93,Jovanovic:JFM05,McKeon:JFM10}. 
	
	We can also incorporate temporal damping into forcing and response modes as $[\hat{\mathbf{q}}_{k_{z^\prime},\omega}, \hat{\mathbf{f}}_{k_{z^\prime},\omega}]e^{-\beta t}$ through a discounted resolvent operator, where $\beta$ is a time-discounting parameter \citep{Jovanovic:04,Schmid:AMR14}. Moreover, the pseudospectral  analysis is dependent on the norm \citep{Trefethen05}. In this work, we use the Chu norm \citep{Chu:Acta65} which is incorporated into the resolvent through a similarity transform $\mathbf{H}_{\bar{\mathbf{q}}} \rightarrow \mathbf{W}^{\frac{1}{2}} \mathbf{H}_{\bar{\mathbf{q}}} \mathbf{W}^{-\frac{1}{2}}$, where $\mathbf{W}$ is the weight matrix  that accounts for numerical quadrature and energy weights.  
	
	The weighted resolvent is dependent on the temporal frequency $\omega$, spanwise wavenumber $k_{z^\prime}$, and the base flow $\bar{\mathbf{q}} = \bar{\mathbf{q}}(\alpha,\Lambda)$. These parameters define a large parameter space to characterize the effect of sweep angle on the wake dynamics. To facilitate this characterization, we employ an adjoint-based parametric sensitivity method for $\omega$ and $k_{z^\prime}$	\citep{Schmid:AMR14,Fosas:JoT17}. This approach is helpful when the parameter space is large as well as to capture the sensitivity of the resolvent norm to specific geometrical and flow parameters \citep{Skene:JFM19}. 
	
	To perform the resolvent analysis, we construct a discrete linear operator $\mathbf{L}_{\bar{\mathbf{q}}}$ \citep{Sun:JFM17}. This operator is discretized over a $2$D unstructured grid, as shown in figure \ref{fig:phys_setup}, with a reduced-size spatial domain $x/L_c \in [-10,15]$ and $y/L_c \in [-10,10]$  to alleviate the computational costs of performing resolvent analysis without affecting the accuracy of the resolvent modes. The structured DNS base flow solution is transferred to the unstructured grid via cubic interpolation.  We enforce homogeneous Dirichlet boundary conditions for the fluctuating variables $\rho'$ and $u'$ and homogeneous Neumann boundary conditions for $T'$ along the farfield and airfoil boundaries as well as to all variables at the computational outlet. In addition, we apply sponges far from airfoil in conjunction with the boundary conditions. 
	
	In the present work, we construct $\mathbf{L}_{\bar{\mathbf{q}}}$ with $m$ ranging between $150,000$ and $200,000$. The resolvent modes were computed using the randomized resolvent algorithm \citep{Ribeiro:PRF20}, sketching the operator with $10$ random test vectors weighted by the gradients of the baseflow $\left(\| \nabla \rho\|, \|\nabla u_x\|, \|\nabla u_y\|, \|\nabla u_z\|, \|\nabla p\|\right)$. The resolvent norm converges to at least $7$ significant digits. For the spectral analysis of $\mathbf{L}_{\bar{\mathbf{q}}}$, eigenmodes were computed using the Krylov--Schur method  \citep{Stewart:SIAM02} with $128$ vectors for the Krylov subspace and tolerance residual of $10^{-10}$. The direct and adjoint linear systems were solved using the \textsf{MUMPS} package.  The codes used to compute the resolvent modes and eigenvalues are part of the `Linear Analysis Package' made available by \cite{SkeneRibeiro:tools}.
	
	\section{Effect of sweep on wake dynamics}
	\label{sec:swept}
	
	\subsection{Wake characterization}
	\label{sec:wake}
	
	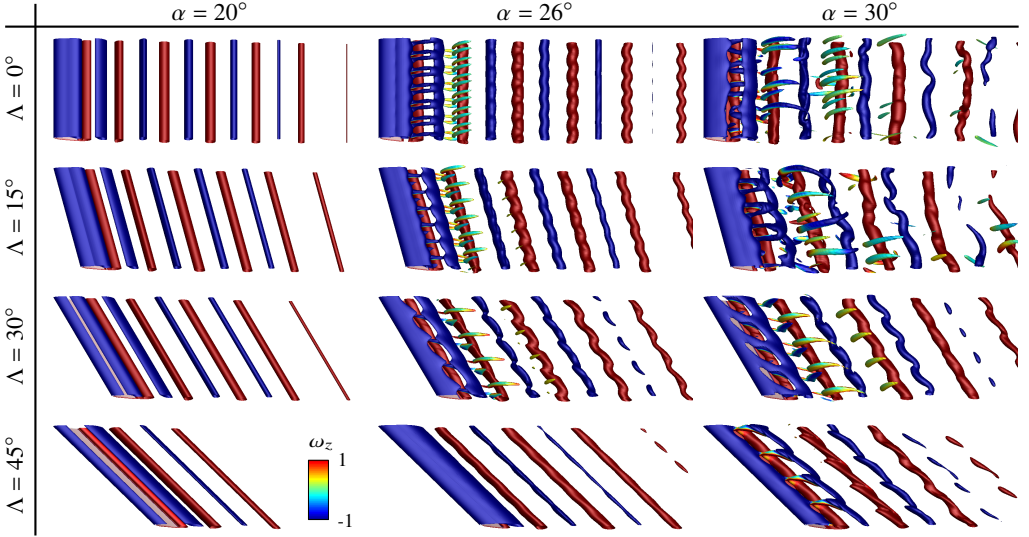
\begin{figure}
		\centering
		\begin{tikzpicture}
		\node[anchor=south west,inner sep=0] (image) at (0.6,0) {\includegraphics[trim=40mm 2mm 30mm 0mm, clip,width=0.31\textwidth]{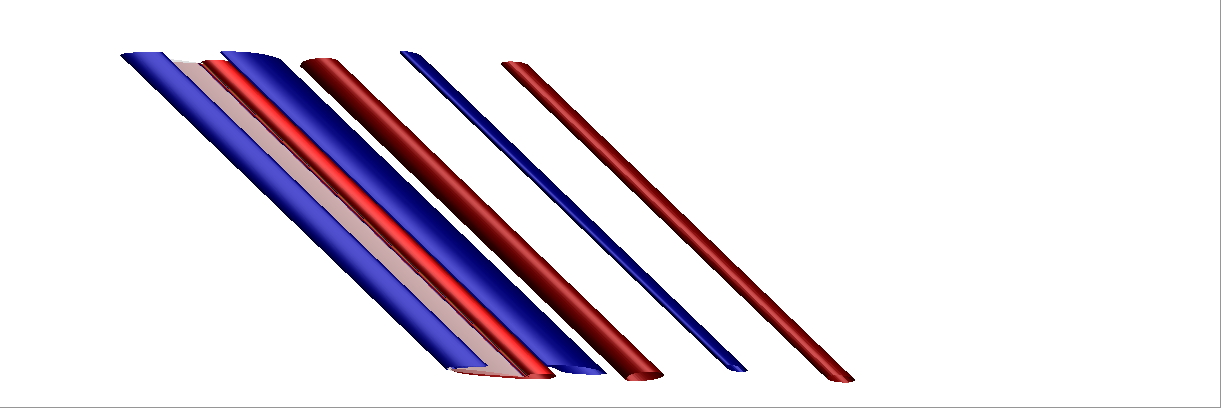}};
		\node[anchor=south west,inner sep=0] (image) at (4.9,0) {\includegraphics[trim=40mm 2mm 30mm 0mm, clip,width=0.31\textwidth]{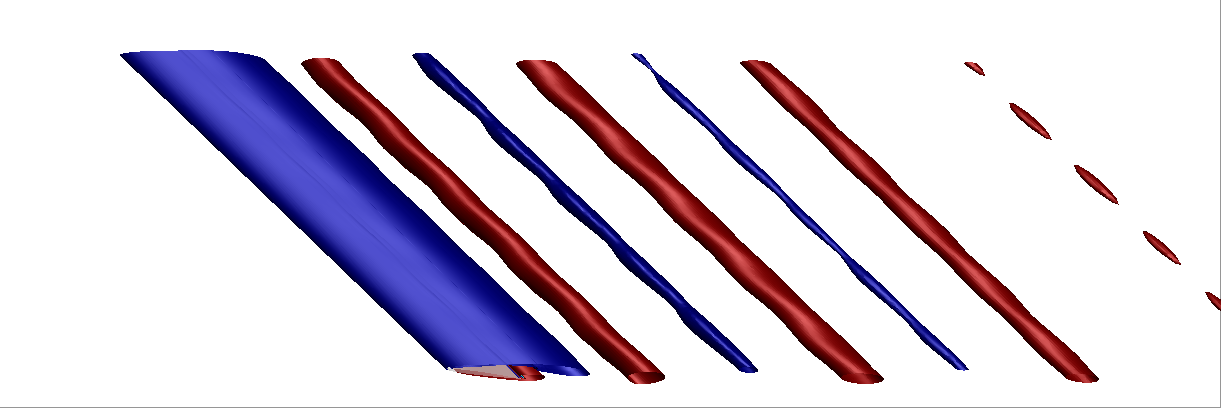}};
		\node[anchor=south west,inner sep=0] (image) at (9.2,0) {\includegraphics[trim=40mm 2mm 30mm 0mm, clip,width=0.31\textwidth]{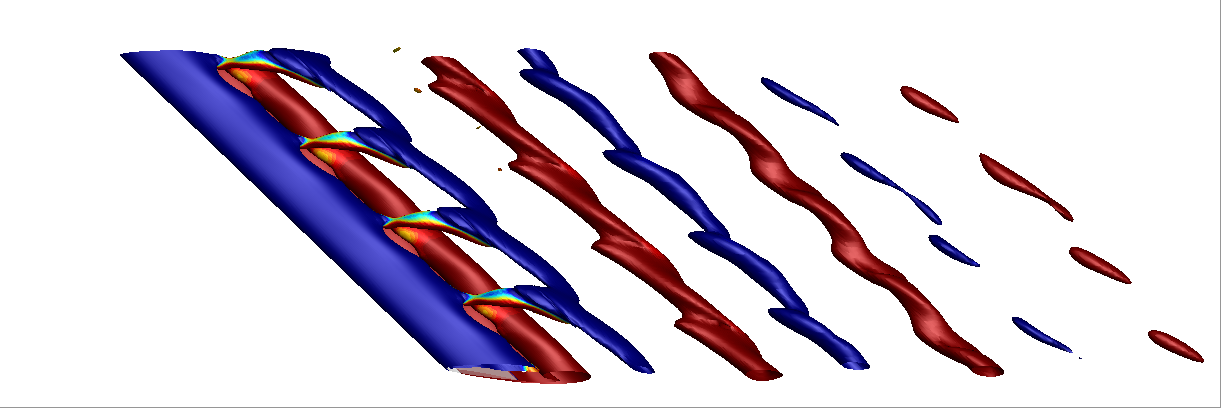}};
		\node[anchor=south west,inner sep=0] (image) at (0.6,1.7) {\includegraphics[trim=40mm 2mm 30mm 0mm, clip,width=0.31\textwidth]{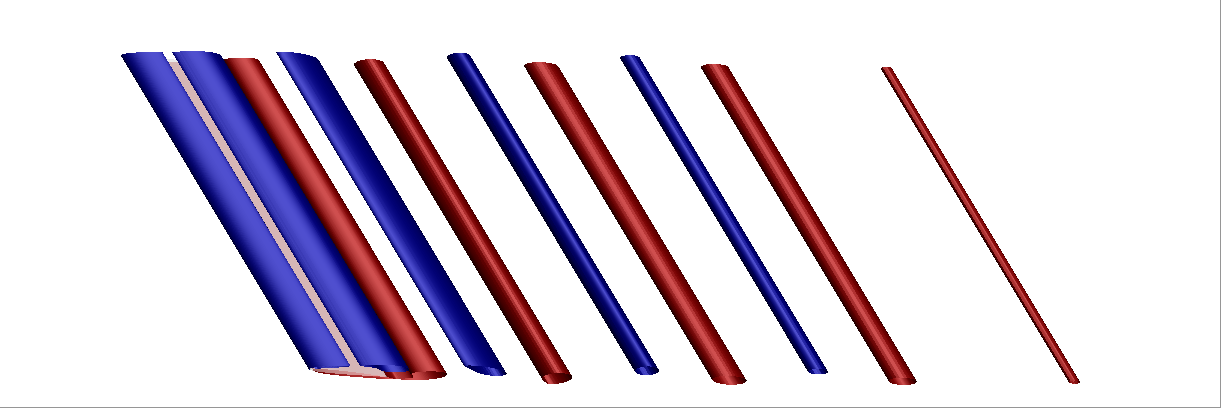}};
		\node[anchor=south west,inner sep=0] (image) at (4.9,1.7) {\includegraphics[trim=40mm 2mm 30mm 0mm, clip,width=0.31\textwidth]{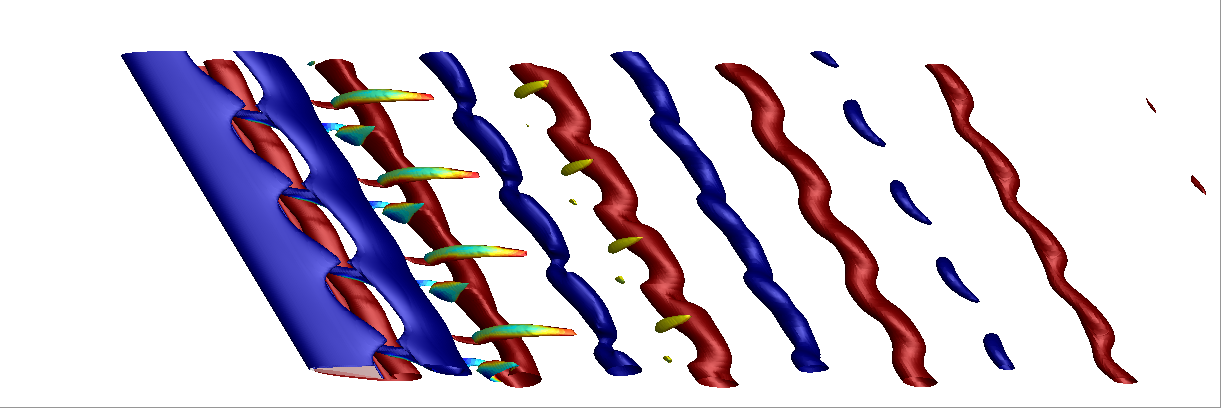}};
		\node[anchor=south west,inner sep=0] (image) at (9.2,1.7) {\includegraphics[trim=40mm 2mm 30mm 0mm, clip,width=0.31\textwidth]{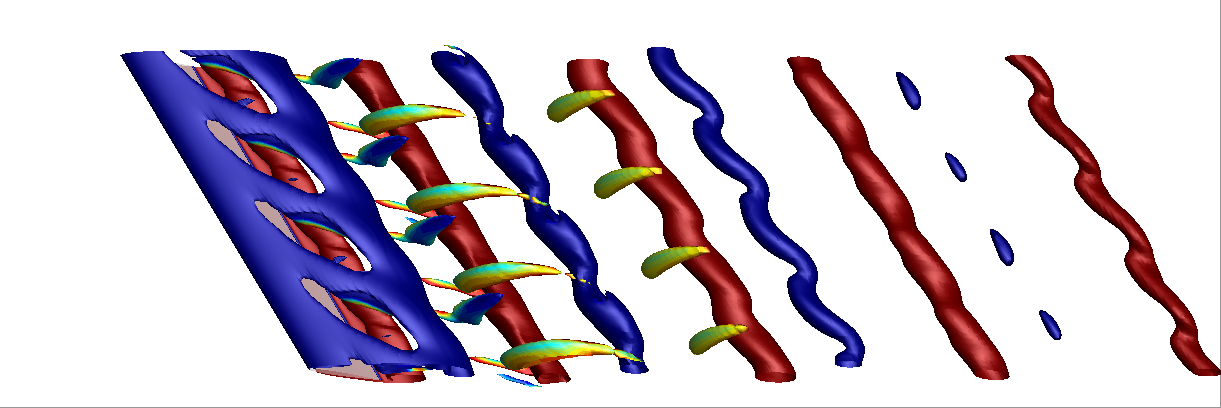}};
		\node[anchor=south west,inner sep=0] (image) at (0.6,3.4) {\includegraphics[trim=40mm 2mm 30mm 0mm, clip,width=0.31\textwidth]{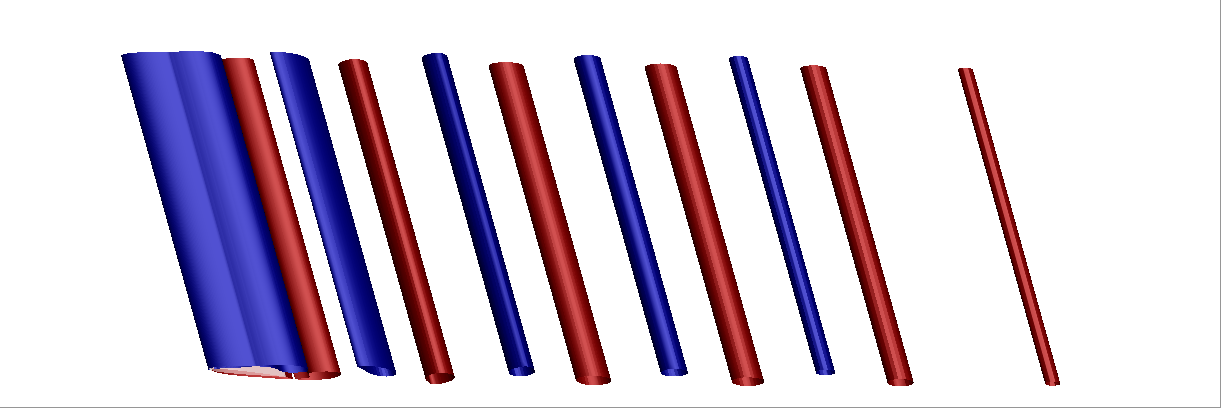}};
		\node[anchor=south west,inner sep=0] (image) at (4.9,3.4) {\includegraphics[trim=40mm 2mm 30mm 0mm, clip,width=0.31\textwidth]{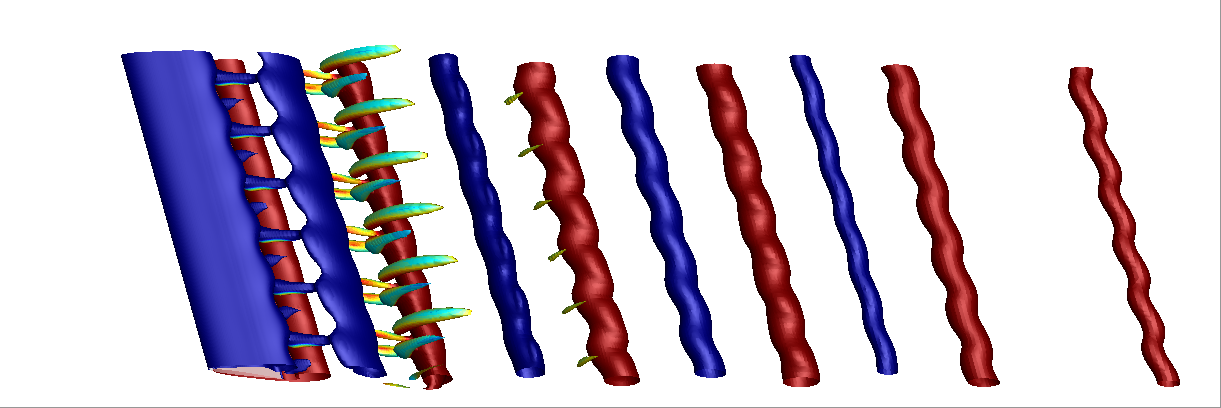}};
		\node[anchor=south west,inner sep=0] (image) at (9.2,3.4) {\includegraphics[trim=40mm 2mm 30mm 0mm, clip,width=0.31\textwidth]{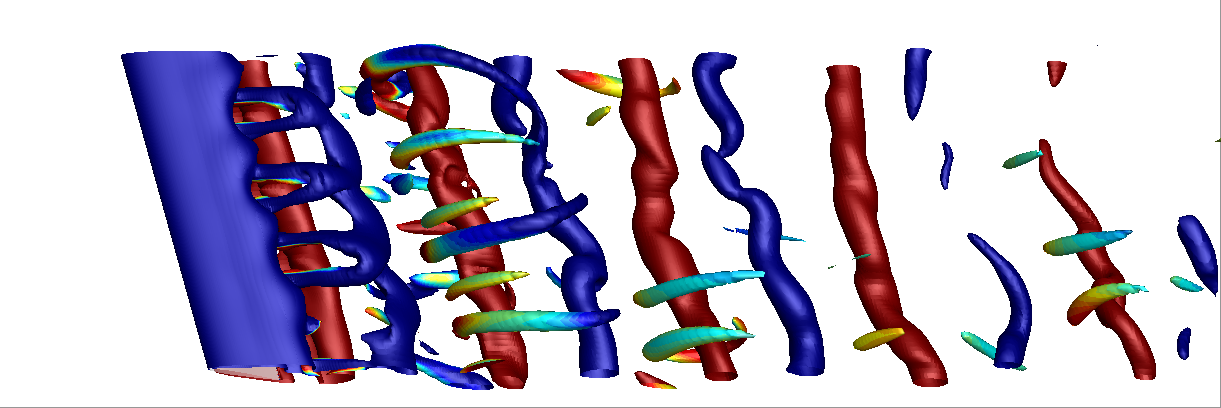}};
		\node[anchor=south west,inner sep=0] (image) at (0.6,5.1) {\includegraphics[trim=40mm 2mm 30mm 0mm, clip,width=0.31\textwidth]{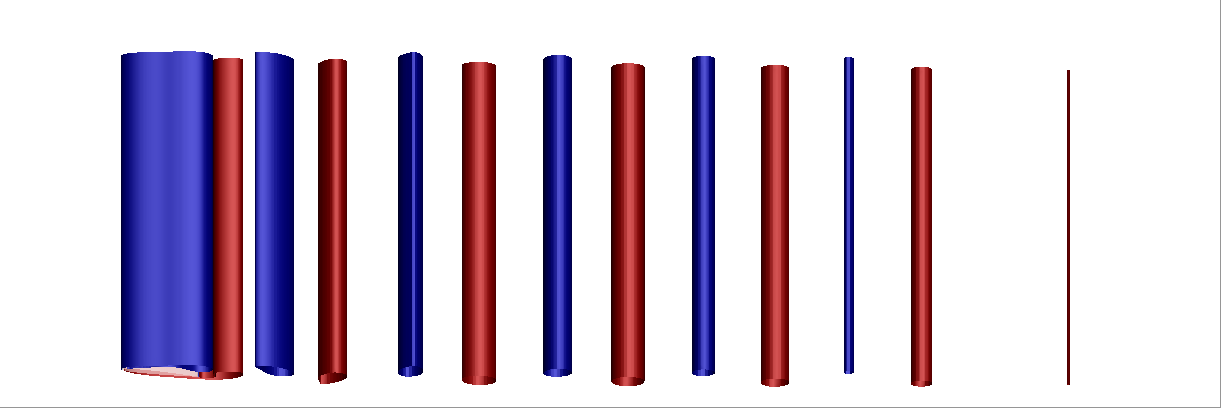}};
		\node[anchor=south west,inner sep=0] (image) at (4.9,5.1) {\includegraphics[trim=40mm 2mm 30mm 0mm, clip,width=0.31\textwidth]{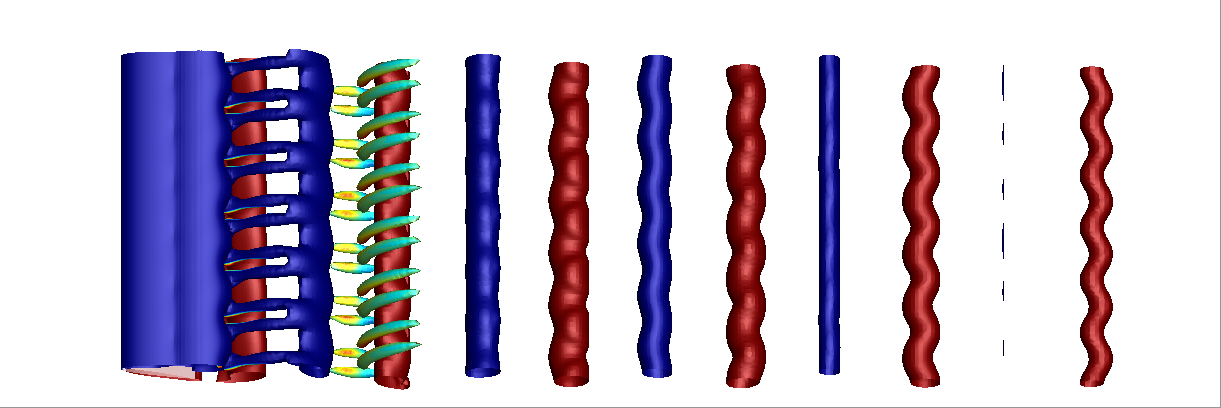}};
		\node[anchor=south west,inner sep=0] (image) at (9.2,5.1) {\includegraphics[trim=40mm 2mm 30mm 0mm, clip,width=0.31\textwidth]{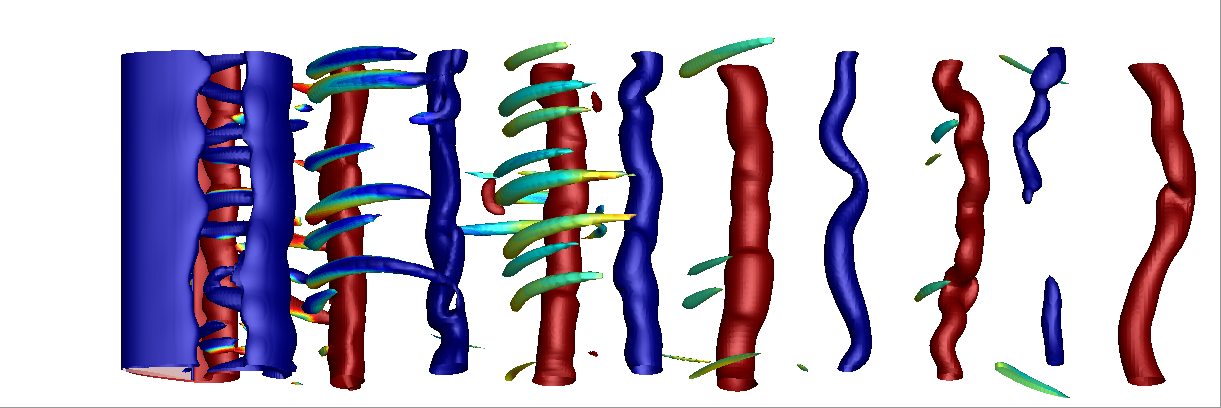}};
		\node[text width=0cm] at (2.2,6.9) {\rotatebox{0}{$\alpha = 20^\circ$}};
		\node[text width=0cm] at (6.5,6.9) {\rotatebox{0}{$\alpha = 26^\circ$}};
		\node[text width=0cm] at (10.8,6.9) {\rotatebox{0}{$\alpha = 30^\circ$}};
		\node[text width=0cm] at (0.0,5.9) {\rotatebox{90}{$\Lambda = 0^\circ$}};
		\node[text width=0cm] at (0.0,4.2) {\rotatebox{90}{$\Lambda = 15^\circ$}};
		\node[text width=0cm] at (0.0,2.5) {\rotatebox{90}{$\Lambda = 30^\circ$}};
		\node[text width=0cm] at (0.0,0.8) {\rotatebox{90}{$\Lambda = 45^\circ$}};
		\draw[black, thick] (0.4,0) -- (0.4,7.0);
		\draw[black, thick] (0,6.7) -- (13.4,6.7);
		\node[anchor=south west,inner sep=0] (image) at (4.0,0.2) {\includegraphics[frame,trim=0mm 0mm 0mm 0mm, clip,width=0.02\textwidth]{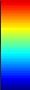}};
		\scriptsize
		\node[text width=0cm] at (4.4,1.0) {\rotatebox{0}{$1$}};
		\node[text width=0cm] at (4.4,0.2) {\rotatebox{0}{-$1$}};
		\node[text width=0cm] at (4.0,1.2) {\rotatebox{0}{$\omega_z$}};
		\end{tikzpicture} \\ \vspace{-2mm} 
		\caption{Instantaneous flow field visualization with isosurfaces of $Q$ criterion colored by the vorticity component $\omega_z$. For $\alpha \ge 26^\circ$, the wake is $3$D. As $\Lambda$ increases, the vortices become slanted with the sweep angle and wake three-dimensionality is reduced for $\alpha \ge 26^\circ$.} 
		\label{fig:inst_qcrit_all}
	\end{figure}
	
	The wake structures are affected by sweep and incidence angles, as shown by the isosurfaces of $Q$ colored by the vorticity $\omega_z$ in figure \ref{fig:inst_qcrit_all}. For angles of attack $\alpha \le 20^\circ$, the flow is $2$D and the wake vortices are aligned with the sweep angle. For $\alpha \ge 26^\circ$, the vortical structures exhibit spanwise oscillations with a transition from $2$D to $3$D vortex shedding. At $\alpha = 26^\circ$, a sinusoidal pattern of oscillations appears in the spanwise direction and, at $\alpha = 30^\circ$, streamwise vortical structures emerge.
	
	Highly swept wings induce spanwise flows and attenuate wake three-dimensionality as evident from the flow visualizations. When the sweep angle is $\Lambda \le 15^\circ$ the wake is similar to the flow over unswept wings for all angles of attack. The wake is significantly altered for sweep angles $\Lambda \ge 30^\circ$, especially at high angles of attack, when spanwise oscillations are advected by the spanwise flow. For instance, at $\Lambda = 45^\circ$ and $\alpha = 26^\circ$, the spanwise oscillations are almost suppressed. The attenuation of spanwise fluctuations also occurs for $\alpha = 30^\circ$, as we observe a similar effect for $\Lambda \ge 30^\circ$. 
	
	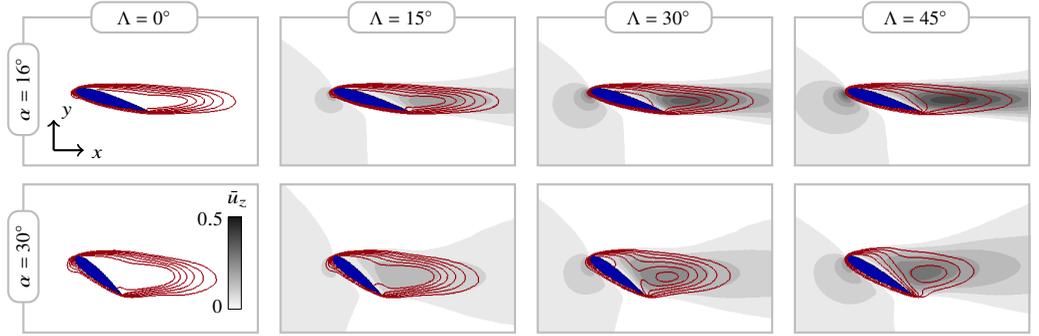
\begin{figure}
		\centering
		\begin{tikzpicture}
		\node[anchor=south west,inner sep=0] (image) at (-0.1,0) {\includegraphics[trim=0mm 2mm 2mm 0mm, clip,width=0.23\textwidth]{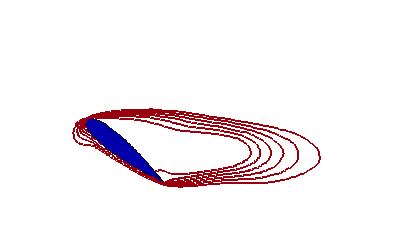}};
		\node[anchor=south west,inner sep=0] (image) at (3.3,0) {\includegraphics[trim=0mm 2mm 2mm 0mm, clip,width=0.23\textwidth]{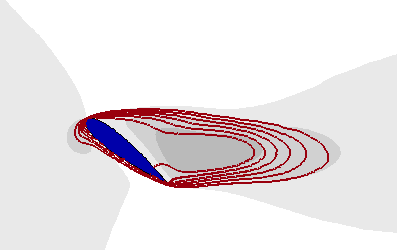}};
		\node[anchor=south west,inner sep=0] (image) at (6.7,0) {\includegraphics[trim=0mm 2mm 2mm 0mm, clip,width=0.23\textwidth]{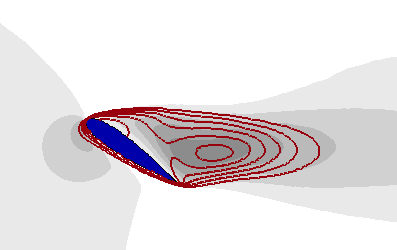}};
		\node[anchor=south west,inner sep=0] (image) at (10.1,0) {\includegraphics[trim=0mm 2mm 2mm 0mm, clip,width=0.23\textwidth]{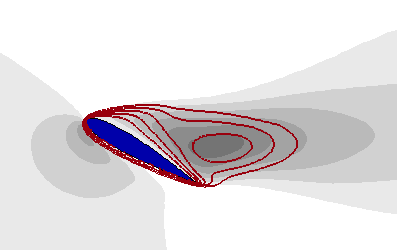}};
		\node[anchor=south west,inner sep=0] (image) at (-0.1,2.2) {\includegraphics[trim=0mm 2mm 2mm 0mm, clip,width=0.23\textwidth]{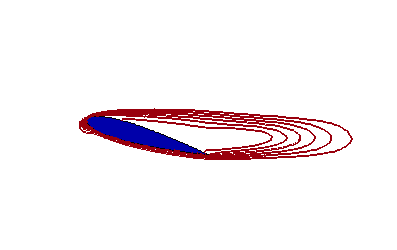}};
		\node[anchor=south west,inner sep=0] (image) at (3.3,2.2) {\includegraphics[trim=0mm 2mm 2mm 0mm, clip,width=0.23\textwidth]{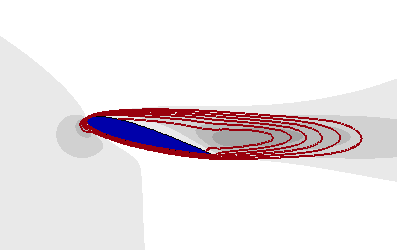}};
		\node[anchor=south west,inner sep=0] (image) at (6.7,2.2) {\includegraphics[trim=0mm 2mm 2mm 0mm, clip,width=0.23\textwidth]{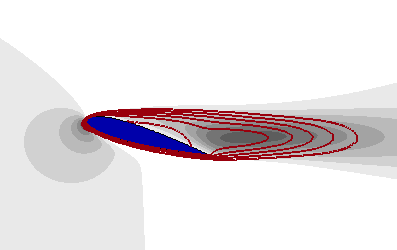}};
		\node[anchor=south west,inner sep=0] (image) at (10.1,2.2) {\includegraphics[trim=0mm 2mm 2mm 0mm, clip,width=0.23\textwidth]{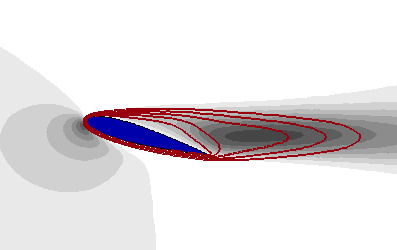}};
		\draw[blueI, thick] (-0.1,0.0) rectangle (3.0,1.95);
		\draw[blueI, thick] (3.3,0.0) rectangle (6.4,1.95);
		\draw[blueI, thick] (6.7,0.0) rectangle (9.8,1.95);
		\draw[blueI, thick] (10.1,0.0) rectangle (13.2,1.95);
		\draw[blueI, thick] (-0.1,2.2) rectangle (3.0,4.15);
		\draw[blueI, thick] (3.3,2.2) rectangle (6.4,4.15);
		\draw[blueI, thick] (6.7,2.2) rectangle (9.8,4.15);
		\draw[blueI, thick] (10.1,2.2) rectangle (13.2,4.15);
		\scriptsize
		\draw[rounded corners, blueI, fill=white, thick] (-0.3,2.6) rectangle (0.1,3.8);
		\node[text width=0cm] at (-0.2,3.2) {\rotatebox{90}{$\alpha = 16^\circ$}};
		\draw[rounded corners, blueI, fill=white, thick] (-0.3,0.4) rectangle (0.1,1.6);
		\node[text width=0cm] at (-0.2,1.0) {\rotatebox{90}{$\alpha = 30^\circ$}};
		\draw[rounded corners, blueI, fill=white, thick] (0.8,3.95) rectangle (2.2,4.35);
		\node[align=left] at (1.5,4.15) {\rotatebox{0}{$\Lambda = 0^\circ$}};
		\draw[rounded corners, blueI, fill=white, thick] (4.2,3.95) rectangle (5.6,4.35);
		\node[align=left] at (4.9,4.15) {\rotatebox{0}{$\Lambda = 15^\circ$}};
		\draw[rounded corners, blueI, fill=white, thick] (7.6,3.95) rectangle (9.0,4.35);
		\node[align=left] at (8.3,4.15) {\rotatebox{0}{$\Lambda = 30^\circ$}};
		\draw[rounded corners, blueI, fill=white, thick] (11.0,3.95) rectangle (12.4,4.35);
		\node[align=left] at (11.7,4.15) {\rotatebox{0}{$\Lambda = 45^\circ$}};
		\node[anchor=south west,inner sep=0] (image) at (2.6,0.3) {\includegraphics[frame,trim=0mm 2mm 2mm 0mm, clip,width=0.013\textwidth,height=0.09\textwidth]{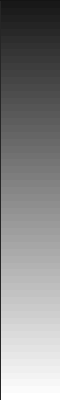}};
		\node[text width=0cm] at (2.6,1.75) {\rotatebox{0}{$\bar{u}_z$}};
		\node[text width=0cm] at (2.20,1.50) {\rotatebox{0}{$0.5$}};
		\node[text width=0cm] at (2.40,0.35) {\rotatebox{0}{$0$}};
		\draw[->,black, thick] (0.30,2.40) -- (0.70,2.40);   
		\draw[->,black, thick] (0.30,2.40) -- (0.30,2.80);   
		\node[text width=0cm] at (0.40,2.90) {$y$};
		\node[text width=0cm] at (0.80,2.35) {$x$};
		\end{tikzpicture} \vspace{-5mm}
		\caption{Time-averaged flow field visualization with $z$--velocity component, $\bar{u}_z \in [0,0.5]$, in grayscale, for $0^\circ \le \Lambda \le 45^\circ$ and $\alpha = 16^\circ$ and $30^\circ$. Red solid contours mark the laminar separation bubble, with $6$ equally distributed isolines of $x$--velocity component, $\bar{u}_x \in [0,0.5]$. 	Crossflow component $\bar{u}_z$ strengthens with $\alpha$ and $\Lambda$. } 
		\label{fig:timeavg_flow}
	\end{figure}
	
	Even though sweep attenuates three-dimensionality, the sustained unsteadiness of the wake suggests the existence of self-supported mechanisms that yield distinct vortex shedding patterns in swept wings at high incidence angles. We gain further insights into the separated flows by analyzing the time-averaged flow field contours of streamwise and spanwise velocities, as seen in figure \ref{fig:timeavg_flow}. In general, as we increase the sweep angle, the $\bar{u}_x = 0$ contour line approaches the wing surface. For $\alpha = 30^\circ$, we also notice a circular $\bar{u}_z$ profile appearing in the wake region where the spanwise flow is stronger.
	
	\begin{figure}
		\centering
		\begin{overpic}[trim=0mm 0mm 0mm 0mm, clip,width=1\textwidth]{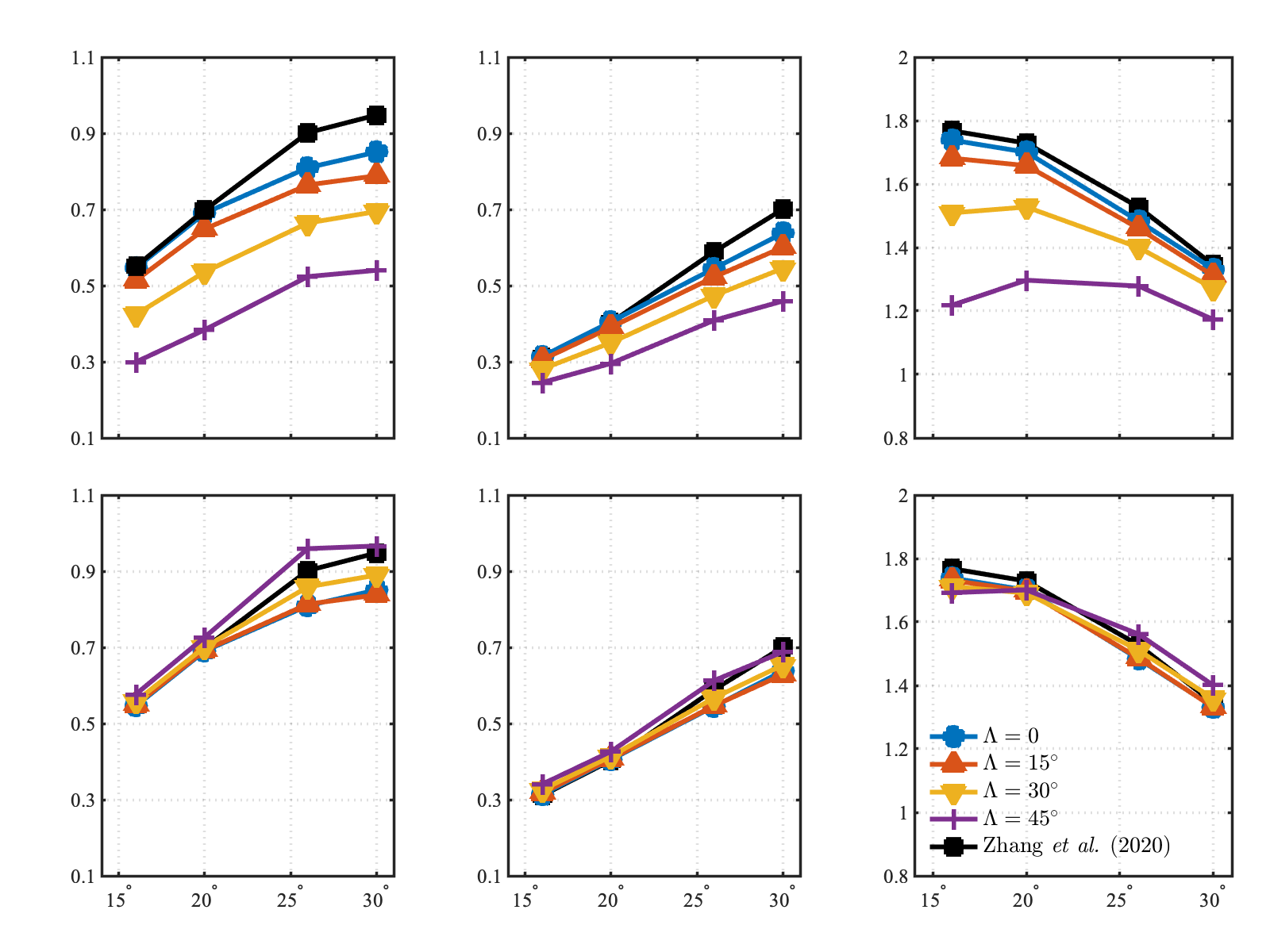}
			\footnotesize
			\put(0,20){\rotatebox{90}{$\overline{C_L^\prime}$}}
			\put(33,20){\rotatebox{90}{$\overline{C_D^\prime}$}}
			\put(65,18){\rotatebox{90}{$\overline{C_L^\prime/C_D^\prime}$}}
			\put(0,53){\rotatebox{90}{$\overline{C_L}$}}
			\put(33,53){\rotatebox{90}{$\overline{C_D}$}}
			\put(65,51){\rotatebox{90}{$\overline{C_L/C_D}$}}
			\put(20,0){\rotatebox{0}{$\alpha$}}
			\put(52,0){\rotatebox{0}{$\alpha$}}
			\put(85,0){\rotatebox{0}{$\alpha$}}
		\end{overpic}
		\caption{Time-averaged lift, $\overline{C_L}$, drag, $\overline{C_D}$, and coefficient ratio $\overline{C_L / C_D}$, for all $\alpha, \Lambda$ pair of the present study compared to $2$D incompressible results shown by \cite{Zhang:JFM20}. The bottom row shows the scaled time-averaged coefficients where the flow is analyzed in ($x',y,z'$), and the results collapse for each $\alpha$, for all sweep angles. } 
		\label{fig:CLCD}
	\end{figure}
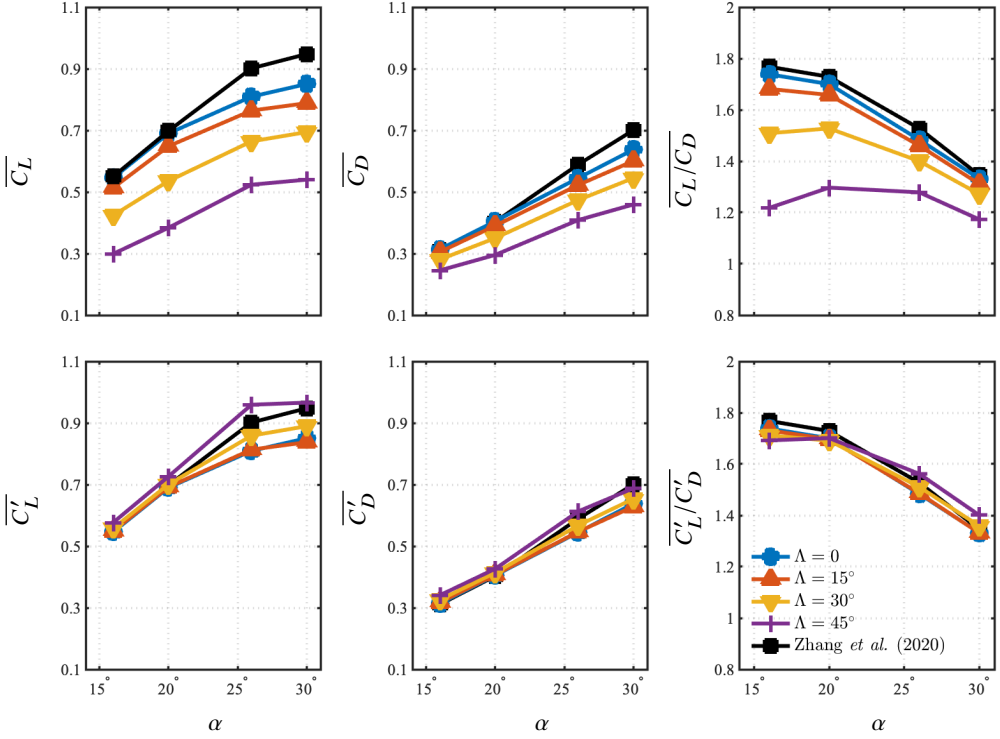
	
	The aerodynamic loads exerted on the wing are also affected by the sweep angle \citep{Zhang:JFM20b,Zhang:PRF22}. However, it is possible to observe similarities among force characteristics with different sweep angles through the independence principle. This leads to the application of proper scaling factors to collapse aerodynamic properties for a variety of  swept wings where adverse pressure gradients and spanwise fluctuations are negligible \citep{Wygnanski:JFM11}.
	
	For the present flows, however, the adverse pressure gradients cannot be neglected due to the  massive separation. In figure \ref{fig:CLCD}, we show that $\overline{C_L}$ and $\overline{C_D}$ differ for the same $\alpha$ if we analyze the flow variables in $(x,y,z)$. The coefficients collapse if we consider scaling the same force coefficients in ($x^\prime,y,z^\prime$). Here, the vector-valued variables in ($x,y,z$) aligned with $x$ are scaled with $\cos\Lambda$ and the effective chord length 
	\begin{equation}
	L_{c}^\prime = L_c  \left( \cos^2 \alpha \cos^2 \Lambda + \sin^2 \alpha \right)^{1/2} \le L_c \mbox{  }
	\label{eq:Lc_prime}
	\end{equation}
	is used to form the scaled $C_L^\prime$ and $C_D^\prime$ coefficients as 
	\begin{equation}
	C_L^\prime = \frac{F_y}{\frac{1}{2}\rho (U_\infty \cos \Lambda)^2 \ L_c^\prime}, \quad C_D^\prime = \frac{F_x \cos\Lambda}{\frac{1}{2}\rho (U_\infty \cos \Lambda)^2 \ L_c^\prime} \mbox{  ,}
	\label{eq:cd_cl_prime}
	\end{equation}
	where $F_x \cos\Lambda = F_{x^\prime}$ is the $x^\prime$ component of the pressure and viscous forces integrated over the airfoil surface per unit depth. As shown in figure \ref{fig:CLCD}(d--f), these scaled coefficients collapse over the angles of attack. Deviations are noticed only for sweep angles $\Lambda \ge 30^\circ$ at high angles of attack $\alpha \ge 26^\circ$. 
	
	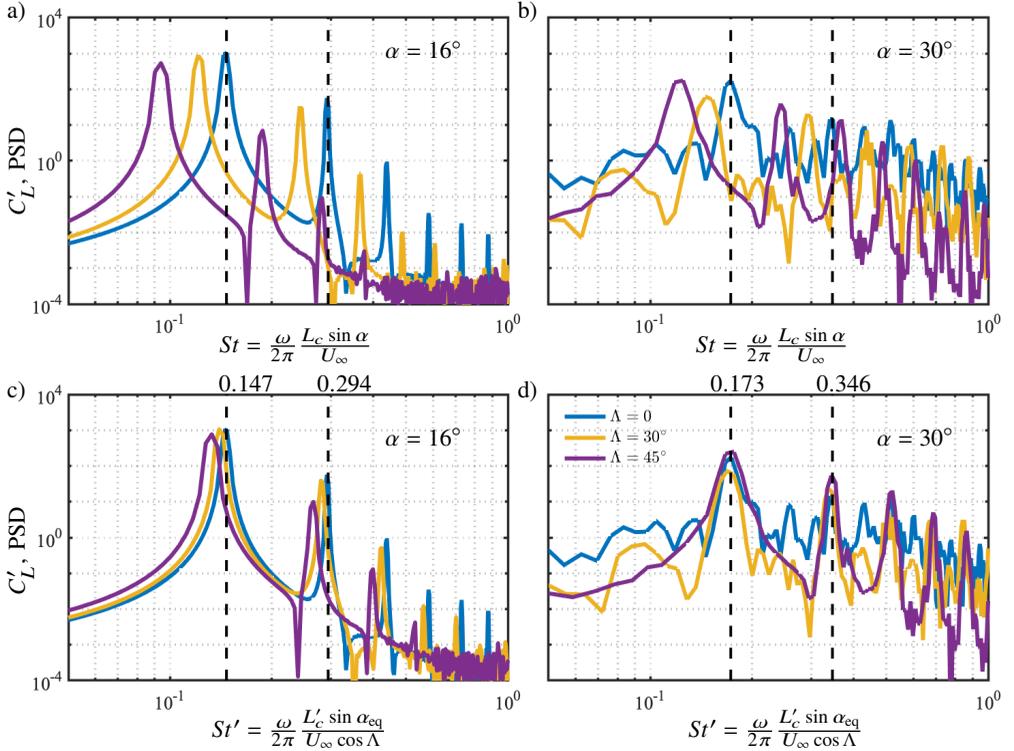
\begin{figure}
		\footnotesize
		\centering
		\begin{tikzpicture}
		\node[anchor=south west,inner sep=0] (image) at (0,0) {\includegraphics[width=1\textwidth]{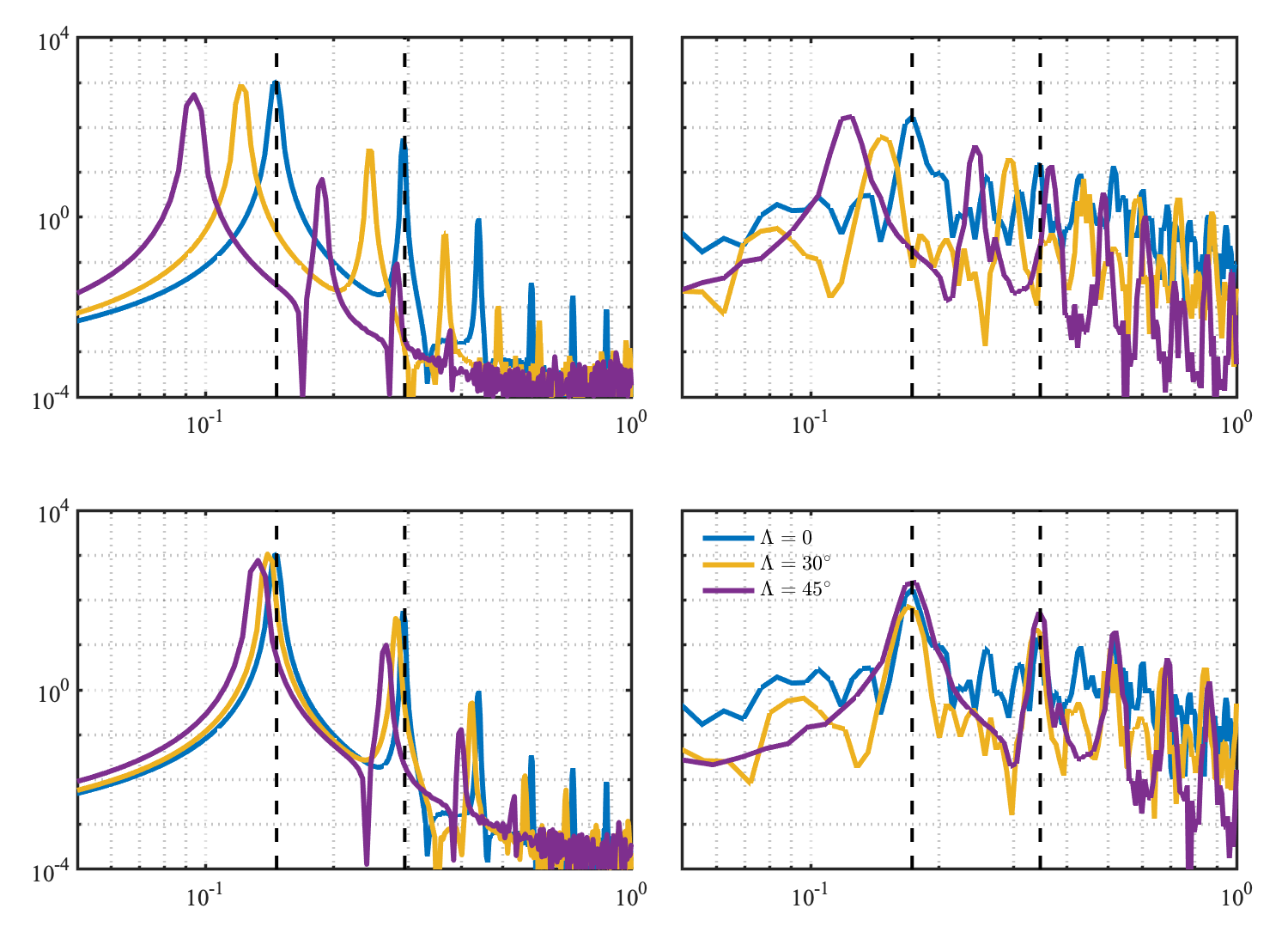}};
		\node[text width=5cm] at (2.5,9.6) {a)};
		\node[text width=5cm] at (9.25,9.6) {b)};
		\node[text width=5cm] at (2.5,4.6) {c)};
		\node[text width=5cm] at (9.25,4.6) {d)};
		\node[text width=5cm] at (7.5,9.1) {$\alpha = 16^\circ$};
		\node[text width=5cm] at (7.5,4.0) {$\alpha = 16^\circ$};
		\node[text width=5cm] at (14.0,9.1) {$\alpha = 30^\circ$};
		\node[text width=5cm] at (14.0,4.0) {$\alpha = 30^\circ$};
		\node[text width=5cm] at (2.5,7.5) {\rotatebox{90}{$C_L^\prime$, PSD}};
		\node[text width=5cm] at (2.5,2.5) {\rotatebox{90}{$C_L^\prime$, PSD}};
		\node[text width=5cm] at (5.3,4.7) {\rotatebox{0}{$0.147$}};
		\node[text width=5cm] at (6.6,4.7) {\rotatebox{0}{$0.294$}};
		\node[text width=5cm] at (11.8,4.7) {\rotatebox{0}{$0.173$}};
		\node[text width=5cm] at (13.2,4.7) {\rotatebox{0}{$0.346$}};
		\node[text width=5cm] at (5.3,5.2) {\rotatebox{0}{$St = \frac{\omega}{2\pi}\frac{L_c \sin \alpha}{U_\infty}$}};
		\node[text width=5cm] at (11.6,5.2) {\rotatebox{0}{$St = \frac{\omega}{2\pi}\frac{L_c \sin \alpha}{U_\infty}$}};
		\node[text width=5cm] at (5.2,0.15) {\rotatebox{0}{$St^\prime = \frac{\omega}{2\pi}\frac{L_c^\prime \sin \alpha_\text{eq}}{U_\infty \cos \Lambda}$}};
		\node[text width=5cm] at (11.5,0.15) {\rotatebox{0}{$St^\prime = \frac{\omega}{2\pi}\frac{L_c^\prime \sin \alpha_\text{eq}}{U_\infty \cos \Lambda}$}};
		\end{tikzpicture}\\ \vspace{-4mm}
		\caption{Scaled lift power spectrum density for (a,c) $\alpha = 16^\circ$ and (b,d) $\alpha = 30^\circ$ at sweep angles $\Lambda = 0^\circ$, $30^\circ$, and $45^\circ$. In (c,d) the Fage--Johansen Strouhal number is analyzed in the $(x',y)$ plane and the dominant and harmonic frequencies for swept wings collapse with the unswept wings.} 
		\label{fig:strouhal}
	\end{figure}
	
	The shown scaling can also be applied to the Fage--Johansen Strouhal number in the $(x^\prime,y)$ plane as 
	\begin{equation}
	St^\prime = \frac{\omega}{2\pi} \frac{L_c^\prime \sin \alpha_\text{eq}}{ U_\infty \cos \Lambda} \mbox{  .}
	\label{eq:st_prime}
	\end{equation}
	As presented in figure \ref{fig:strouhal}, the power spectrum density profiles of $C_L^\prime$ for swept wings exhibit peaks at their characteristic Strouhal number of the vortex shedding and its harmonics.  When the adapted Fage--Johansen Strouhal number $St^\prime$ is considered, the spectral peaks collapse at the same frequencies as observed for the unswept wings. For $\alpha = 16^\circ$, the flow is characterized by a single $2$D vortex shedding and the $C_L^\prime$ spectra is smooth with distinct peak values. For $\alpha = 30^\circ$, the spectra exhibits secondary peaks for $\Lambda \le 30^\circ$ and is smooth for $\Lambda = 45^\circ$, when three-dimensionality is attenuated.
	
\begin{figure}
		\footnotesize
		\centering
		\begin{tikzpicture}
		\node[anchor=south west,inner sep=0] (image) at (0,0) {\includegraphics[width=1\textwidth]{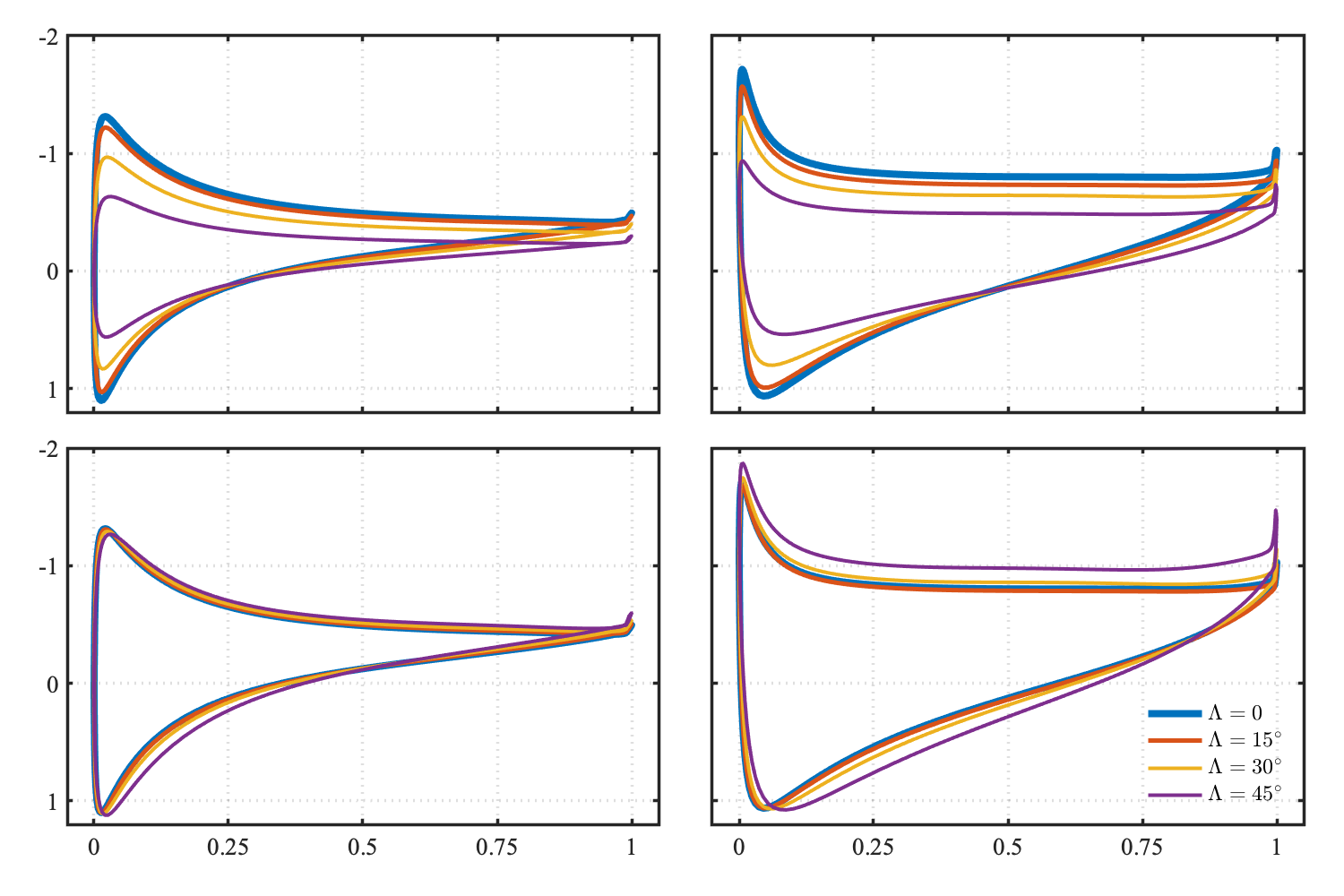}};
		\node[text width=5cm] at (2.5,8.7) {a)};
		\node[text width=5cm] at (9.25,8.7) {b)};
		\node[text width=5cm] at (2.5,4.6) {c)};
		\node[text width=5cm] at (9.25,4.6) {d)};
		\node[text width=5cm] at (4.5,5.6) {pressure side};
		\node[text width=5cm] at (4.5,7.6) {suction side};
		\node[text width=5cm] at (7.5,8.3) {$\alpha = 16^\circ$};
		\node[text width=5cm] at (7.5,4.1) {$\alpha = 16^\circ$};
		\node[text width=5cm] at (14.0,8.3) {$\alpha = 30^\circ$};
		\node[text width=5cm] at (14.0,4.1) {$\alpha = 30^\circ$};
		\node[text width=5cm] at (2.5,6.9) {\rotatebox{90}{$C_p$}};
		\node[text width=5cm] at (2.5,2.7) {\rotatebox{90}{$C_p / \cos^2 \Lambda$}};
		\node[text width=5cm] at (6.0,0.15) {\rotatebox{0}{$x/L_c$}};
		\node[text width=5cm] at (12.4,0.15) {\rotatebox{0}{$x/L_c$}};
		\end{tikzpicture}\\ \vspace{-4mm}
		\caption{Time-averaged pressure coefficients top $C_p$  and bottom $C_p/\cos^2 \Lambda$ over the airfoil surface at (a,c) $\alpha = 16^\circ$ and (b,d) $30^\circ$. } 
		\label{fig:timeavg_cp}
	\end{figure}
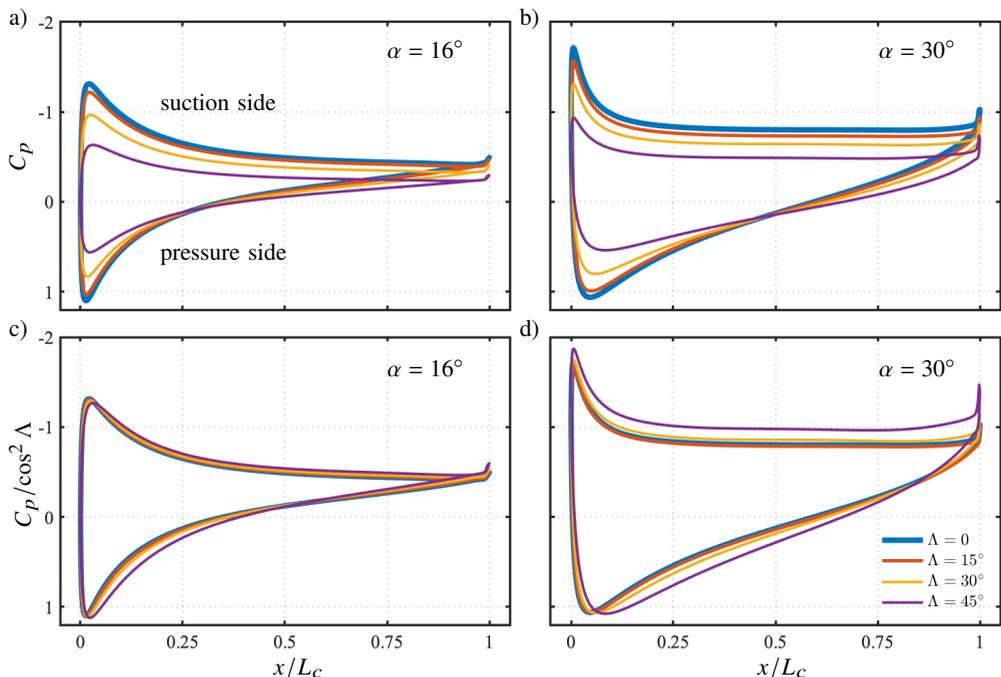

	We can also similarly normalize the pressure coefficients $C_p$ by considering the $U_\infty  \cos\Lambda$ in place of $U_\infty$ in equation \ref{eq:cd_cl}. Indeed, large differences in $C_p$ distribution over the wing are shown in figure \ref{fig:timeavg_cp}(a,b), however, if we consider the scaled-$C_p$, we reveal that the pressure distributions collapse for moderate angles of attack, even though these flows exhibit massive separation, as shown in figure \ref{fig:timeavg_cp}(c,d).	Although we can bring pressure coefficients closer using $C_p / \cos^2 \Lambda$, we notice some deviations for higher angles of sweep and attack, as observed for $\Lambda = 45^\circ$ and $\alpha = 30^\circ$. This motivates us to further understand how massively separated streamwise flow and the strong spanwise flow may impose additional loads over the wing. These deviations indeed suggest that even when the three-dimensionality is attenuated, the wake over laminar swept wings can exert additional aerodynamic forces over the wing for $\Lambda \ge 30^\circ$ and $\alpha \ge 26^\circ$. 
	
	\subsection{Force element analysis}
	\label{sec:forceelements}
	To further understand the sources of deviations in the independence principle, we use the force element theory of \cite{Chang:PRSA92} to relate the near-body vortical structures to aerodynamic forces. Through this method, we identify lift and drag force elements in the near-wake region of the current low-Reynolds number flows and analyze the distribution of force elements near the surface as the wing is swept. This analysis captures the emerging wake structures over swept wings at high angles of attack that exert nonlinear post-stall forces onto the wing.
	
	We observe that the emergence of these force elements is associated with a departure from the collapsed force and pressure coefficients. The sweep-angle dependent scaling in equations \ref{eq:cd_cl_prime} and \ref{eq:st_prime}  assumes independence of streamwise and spanwise flow components. The interaction between them can cause a departure from the collapsed force and pressure coefficients. By using the force element analysis, we uncover near-wake structures that are responsible for the extra forces at high angles of sweep and attack.
	
	
	
	To perform this analysis, we start by defining an auxiliary potential with a specific boundary condition of $-\mathbf{n} \cdot \nabla \phi_i = \mathbf{n} \cdot \mathbf{e}_i$ along the wing surface, where $\phi$ is the auxiliary potential, $\mathbf{n}$ is the unit wall normal vector, and $\mathbf{e}_i$ is normal vectors in the $i$th--direction. For a solenoidal velocity field, the force exerted on the wing in the $i$-th direction can be written as 
	\begin{equation}
	F_i = -\int_V \mathbf{u} \times \bs{\omega} \cdot \nabla \phi_i \text{d}V + \frac{1}{Re} \int_S \mathbf{n} \times \bs{\omega} \cdot (\nabla \phi_i + \mathbf{e_i}) \text{d}S,
	\end{equation}
	where the first integral term on the right-hand side is named vortical elements force and the second integral term is called surface element force. We can visualize the lift and drag force elements by the Hadamard product of the  $\nabla \phi_i$ and the Lamb vector ($\bs{\omega} \times \mathbf{u}$). The auxiliary potential field decays rapidly far from the wing surface, hence the force elements of $(\mathbf{u} \times \bs{\omega}) \circ \nabla \phi_i$  are concentrated near the wing. 
	
	\begin{figure}
		\centering
		\begin{tikzpicture}
		\node[anchor=south west,inner sep=0] (image) at (0.0,0.0) {\includegraphics[trim=0mm 0mm 0mm 0mm, clip,width=0.48\textwidth]{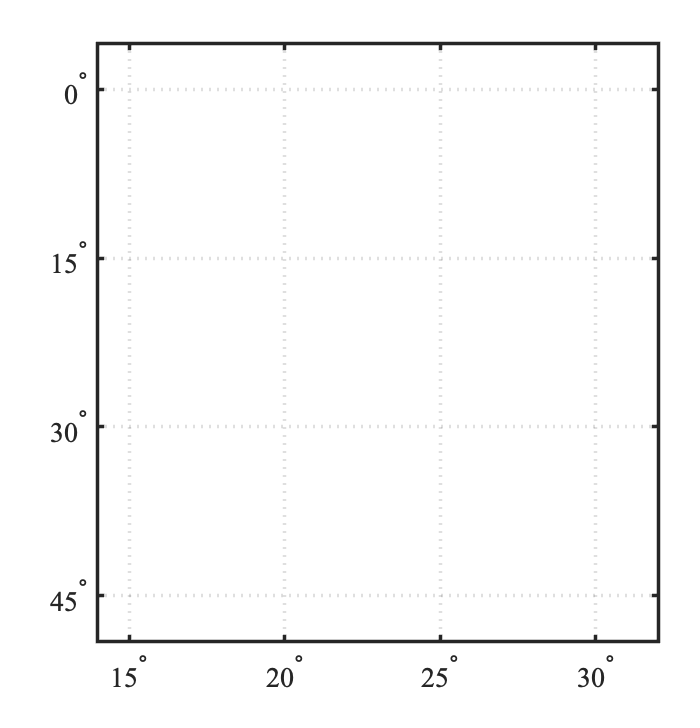}};
		\node[anchor=south west,inner sep=0] (image) at (6.2,4.5) {\includegraphics[trim=15mm 2mm 2mm 25mm, clip,width=0.25\textwidth]{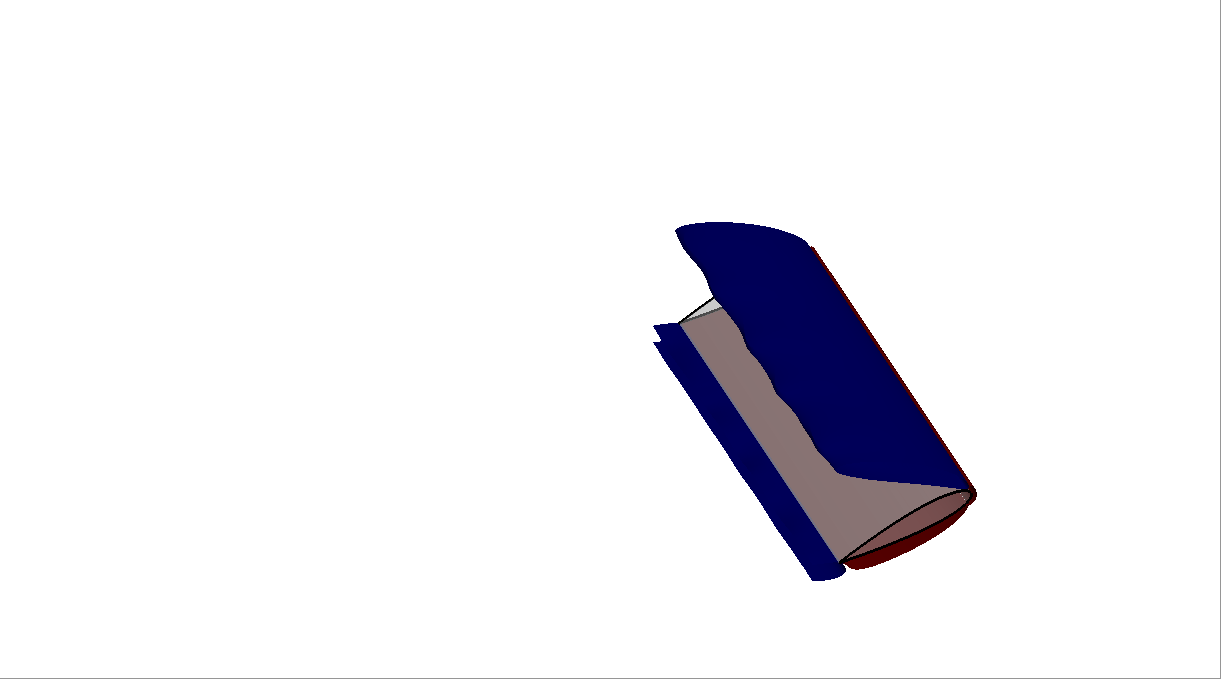}};
		\node[anchor=south west,inner sep=0] (image) at (6.2,2.6) {\includegraphics[trim=15mm 2mm 2mm 25mm, clip,width=0.25\textwidth]{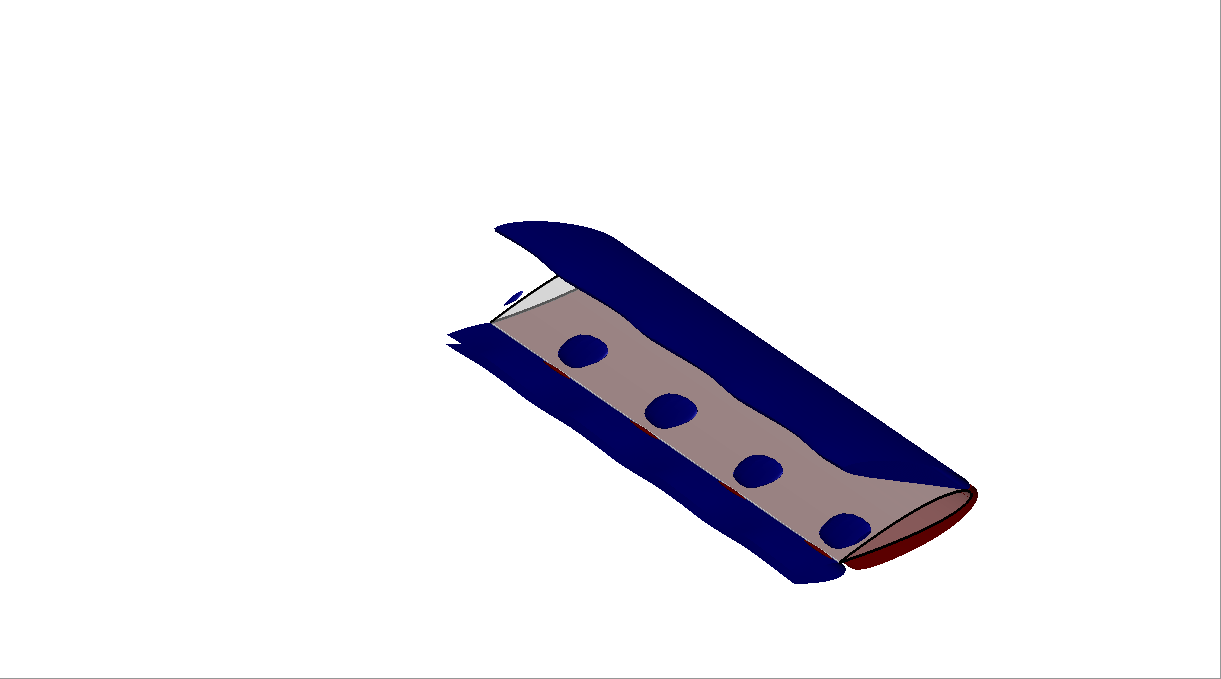}};
		\node[anchor=south west,inner sep=0] (image) at (6.2,0.7) {\includegraphics[trim=15mm 2mm 2mm 25mm, clip,width=0.25\textwidth]{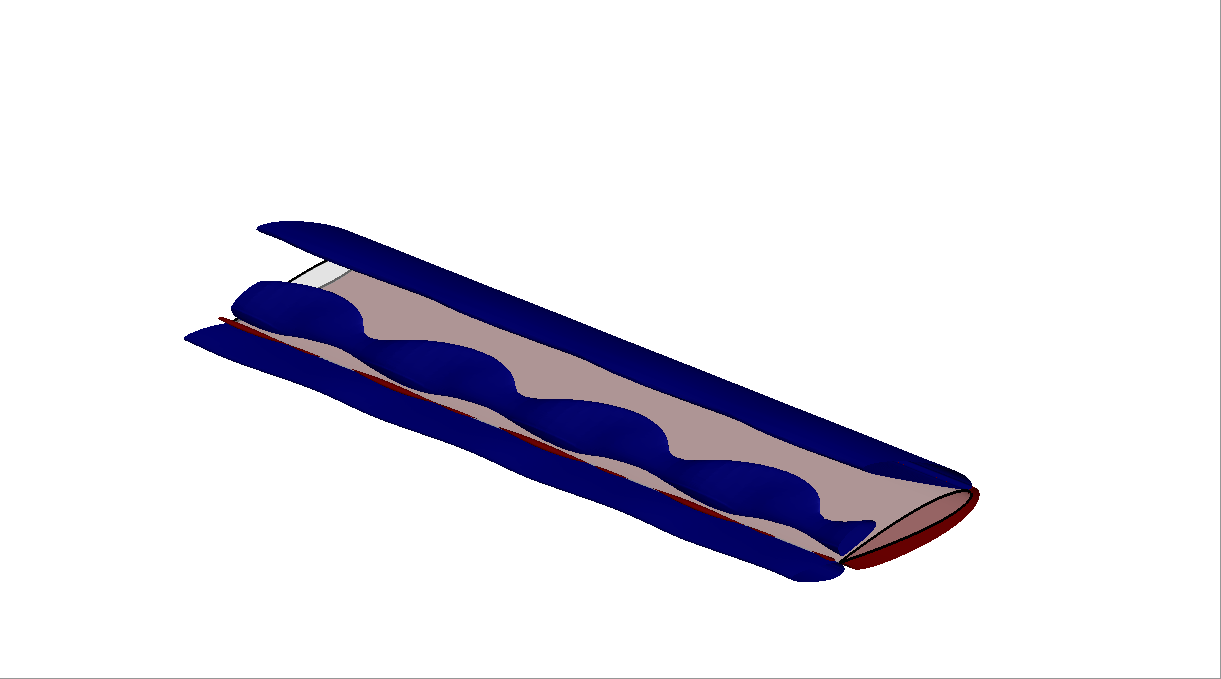}};
		\node[anchor=south west,inner sep=0] (image) at (9.8,4.5) {\includegraphics[trim=0mm 2mm 2mm 10mm, clip,width=0.27\textwidth]{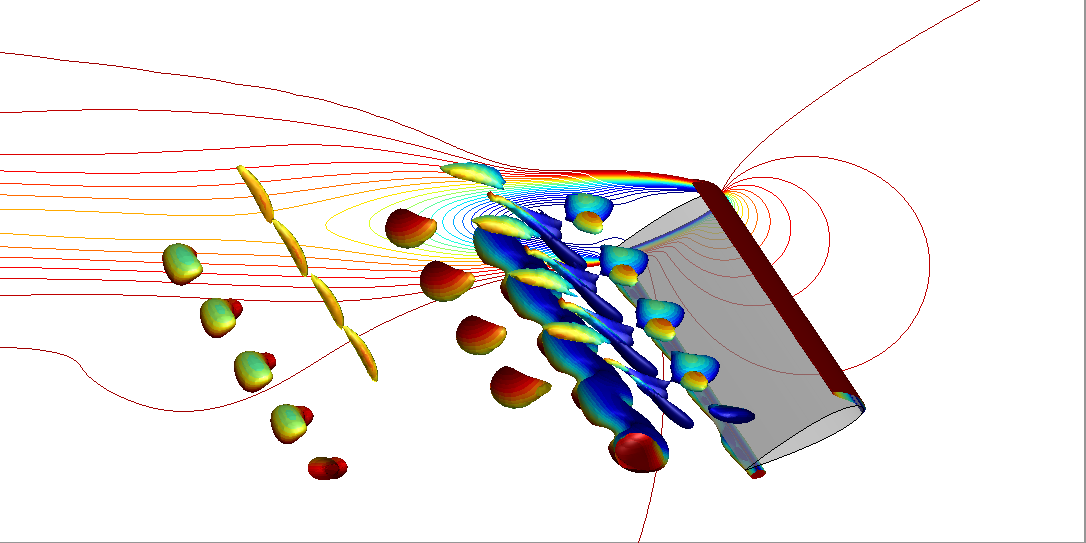}};
		\node[anchor=south west,inner sep=0] (image) at (9.8,2.6) {\includegraphics[trim=0mm 2mm 2mm 10mm, clip,width=0.27\textwidth]{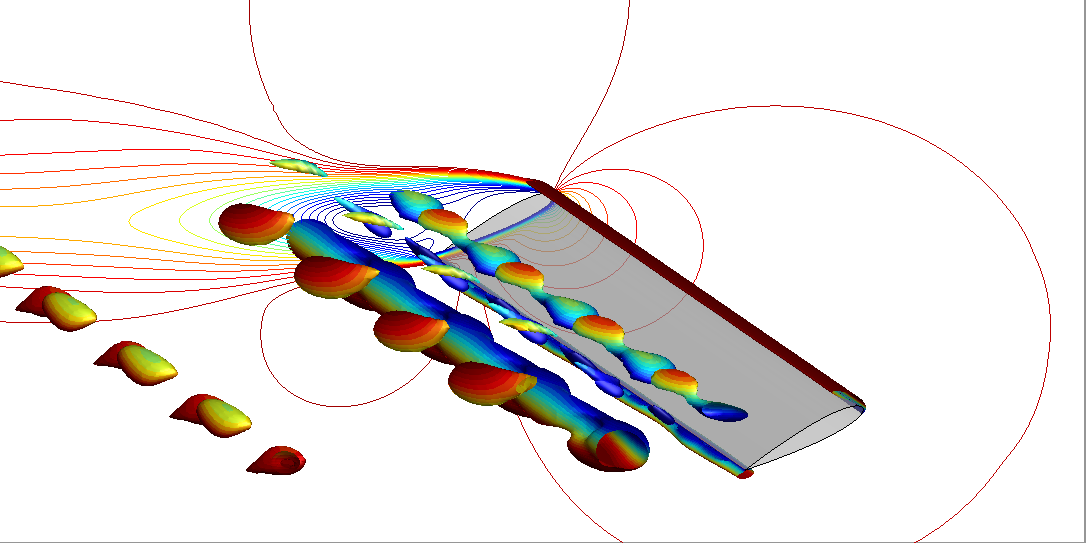}};
		\node[anchor=south west,inner sep=0] (image) at (9.8,0.7) {\includegraphics[trim=0mm 2mm 2mm 10mm, clip,width=0.27\textwidth]{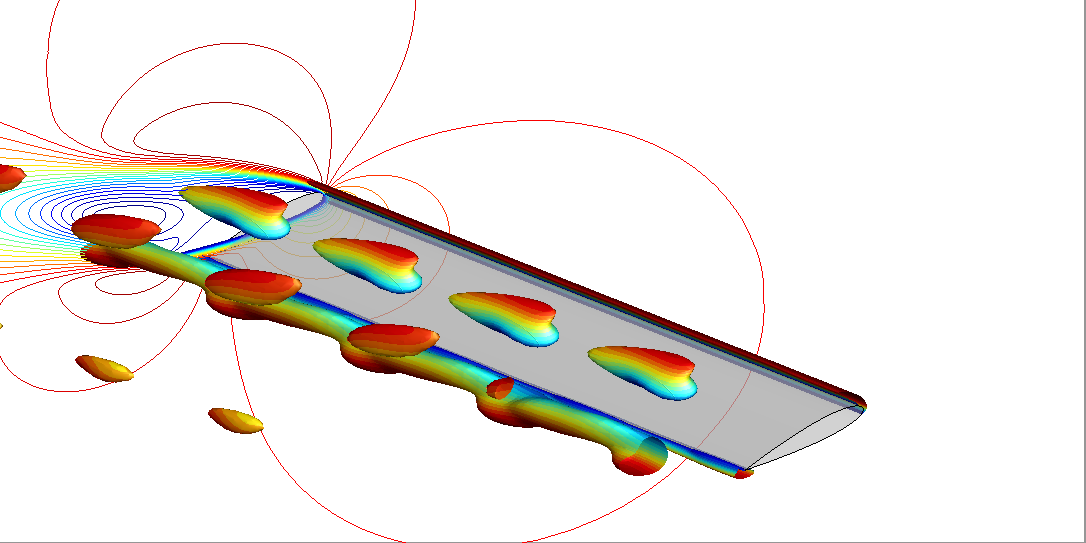}};
		\node[anchor=south west,inner sep=0] (image) at (12.9,3.2) {\includegraphics[frame,trim=0mm 0mm 0mm 0mm, clip,width=0.015\textwidth]{figs/JFM_forceElement_flow_cb.png}};
		\draw[rounded corners,yellow2, very thick] (6.2,4.5) rectangle (9.6,6.2);
		\draw[rounded corners,yellow2, very thick] (9.8,4.5) rectangle (13.45,6.2);
		\draw[<-,yellow2, thick] (6.05,5.2) .. controls (5.5,5.2) .. (5.5,4.42);
		\draw[rounded corners,blue2, very thick] (6.2,2.6) rectangle (9.6,4.3);
		\draw[rounded corners,blue2, very thick] (9.8,2.6) rectangle (13.45,4.3);
		\draw[<-,blue2, thick] (6.05,3.5) .. controls (5.52,3.5) .. (5.52,2.87);
		\draw[rounded corners,dkred2, very thick] (6.2,0.7) rectangle (9.6,2.4);
		\draw[rounded corners,dkred2, very thick] (9.8,0.7) rectangle (13.45,2.4);
		\draw[<-,dkred2, thick] (6.05,1.9) .. controls (5.5,1.9) .. (5.5,1.30);
		\scriptsize
		\node[text width=0cm] at (7.2,6.4) {\rotatebox{0}{force elements}};
		\node[text width=0cm] at (10.8,6.4) {\rotatebox{0}{$Q$-criterion}};
		\huge
		\node[text width=1.5cm] at (1.90,5.85) {\color{yellow2}$\bs{\circ}$};
		\node[text width=1.5cm] at (3.19,5.85) {\color{yellow2}$\bs{\circ}$};
		\node[text width=1.5cm] at (4.77,5.85) {\color{yellow2}$\bs{\circ}$};
		\node[text width=1.5cm] at (6.07,5.85) {\color{yellow2}$\bs{\circ}$};
		\node[text width=1.5cm] at (1.90,4.30) {\color{yellow2}$\bs{\circ}$};
		\node[text width=1.5cm] at (3.19,4.30) {\color{yellow2}$\bs{\circ}$};
		\node[text width=1.5cm] at (4.77,4.30) {\color{yellow2}$\bs{\circ}$};
		\node[text width=1.5cm] at (6.07,4.30) {\color{yellow2}$\bs{\circ}$};
		\node[text width=1.5cm] at (1.90,2.73) {\color{yellow2}$\bs{\circ}$};
		\node[text width=1.5cm] at (3.19,2.73) {\color{yellow2}$\bs{\circ}$};
		\node[text width=1.5cm] at (1.90,1.18) {\color{yellow2}$\bs{\circ}$};
		\node[text width=1.5cm] at (3.19,1.18) {\color{yellow2}$\bs{\circ}$};
		\Large
		\node[text width=1.5cm] at (4.82,2.73) {\color{blue2}$\bs{\square}$};
		\node[text width=1.5cm] at (6.11,2.73) {\color{blue2}$\bs{\square}$};
		\node[text width=1.5cm] at (4.80,1.18) {\color{dkred2}$\bs{\triangle}$};
		\node[text width=1.5cm] at (6.07,1.18) {\color{dkred2}$\bs{\triangle}$};
		\small
		\node[align=left] at (3.5,0.2) {$\alpha$};
		\node[align=left] at (0.2,3.4) {\rotatebox{90}{$\Lambda$}};
		\tiny
		\node[text width=0cm] at (12.4,4.0) {\rotatebox{0}{$\bar{u}_x,u_x$}};
		\node[text width=0cm] at (13.2,3.1) {\rotatebox{0}{$0$}};
		\node[text width=0cm] at (13.2,3.8) {\rotatebox{0}{$1$}};
		\end{tikzpicture} \vspace{-7mm}
		\caption{Characterization of force elements on swept wings. The symbols refer to: {\large \color{yellow2}$\bs{\circ}$} force elements only near the leading and trailing edges, {\color{blue2}$\bs{\square}$} additional equally-spaced small lift elements, and {\color{dkred2}$\bs{\triangle}$} large force structures observed on the suction side. On the right, we show isosurfaces of lift force elements, $(\bs{\omega} \times \mathbf{u}) \cdot \nabla \phi_L \in [-0.3,0.3]$ and  isosurfaces of $Q$ values colored by streamwise velocity component $\bar{u}_x$ with range $[0,1]$ for the time step with the highest lift coefficient $C_L^\prime$ for $\alpha = 30^\circ$ and $15^\circ \le \Lambda \le 45^\circ$. } 
		\label{fig:forceelements}
	\end{figure}
	
	Sweep has a strong influence in limiting the validity of the independence principle at high angles of incidence and it favors the emergence of additional force elements near the wing surface. To show this, let us reveal the vortical structures that generate lift $(\bs{\omega} \times \mathbf{u}) \cdot \nabla \phi_y$ at the instance when the maximum lift is achieved, as shown in figure \ref{fig:forceelements}. Drag elements have similar behavior as the lift and are not shown for brevity. For $\alpha \le 20^\circ$, the force elements are located near the airfoils leading and trailing edges, in the shear-dominated region of the flow, along the edge of the laminar separation bubble.
	
	Additional lift elements appear for higher angles of sweep and attack, as shown in figure \ref{fig:forceelements}. These lift elements are observed over the final quarter chord of the airfoil on the suction side, and as the sweep angle increases, they also increase in size. The force elements can be contrasted with the vortical structures in $Q$ criterion visualization in figure \ref{fig:forceelements}. As such coherent structures are present inside the laminar separation bubble, with size and shape similar to the force elements they can be identified as the lift elements related to the larger deviations in figure \ref{fig:timeavg_cp}(c,d). 
	
	We observe that sweep affects the coherent structures, time-averaged flow fields, and aerodynamic loads and, although some similarities are perceived, sweep has a strong influence on the wake flow downstream at the higher angles of attack. This suggests that flow perturbations originated near the airfoil in the laminar flows over swept wings can be related to the features observed in the nonlinear simulations. To further understand how sweep alters the vortex dynamics we conduct global resolvent analysis to identify the sources of self-sustained mechanisms near the wing that affect the wake behavior.
	
	\subsection{Resolvent analysis}
	\label{sec:resolvent}
	
\begin{figure}
		\centering
		\begin{tikzpicture}
		\node[anchor=south west,inner sep=0] (image) at (-0.1,0) {\includegraphics[trim=0mm 2mm 2mm 0mm, clip,width=0.21\textwidth]{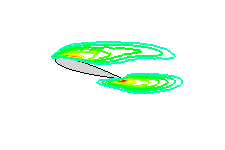}};
		\node[anchor=south west,inner sep=0] (image) at (3.3,0) {\includegraphics[trim=0mm 2mm 2mm 0mm, clip,width=0.21\textwidth]{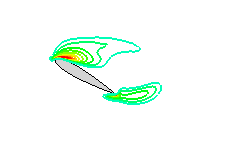}};
		\node[anchor=south west,inner sep=0] (image) at (6.7,0) {\includegraphics[trim=0mm 2mm 2mm 0mm, clip,width=0.21\textwidth]{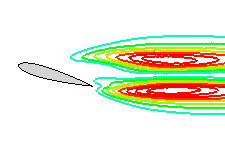}};
		\node[anchor=south west,inner sep=0] (image) at (10.1,0) {\includegraphics[trim=0mm 2mm 2mm 0mm, clip,width=0.21\textwidth]{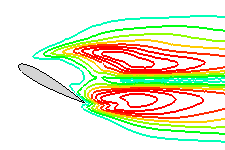}};
		\node[anchor=south west,inner sep=0] (image) at (-0.1,2.2) {\includegraphics[trim=0mm 2mm 2mm 0mm, clip,width=0.21\textwidth]{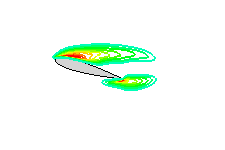}};
		\node[anchor=south west,inner sep=0] (image) at (3.3,2.2) {\includegraphics[trim=0mm 2mm 2mm 0mm, clip,width=0.21\textwidth]{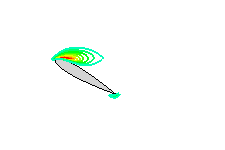}};
		\node[anchor=south west,inner sep=0] (image) at (6.7,2.2) {\includegraphics[trim=0mm 2mm 2mm 0mm, clip,width=0.21\textwidth]{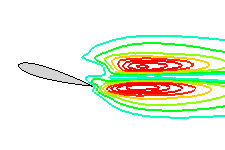}};
		\node[anchor=south west,inner sep=0] (image) at (10.1,2.2) {\includegraphics[trim=0mm 2mm 2mm 0mm, clip,width=0.21\textwidth]{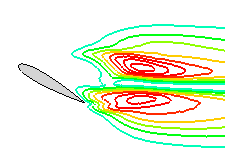}};
		\draw[rounded corners,red2, ultra thick] (-0.1,0.0) rectangle (2.8,1.9);
		\draw[rounded corners,red2, ultra thick] (3.3,0.0) rectangle (6.2,1.9);
		\draw[rounded corners,blue2, ultra thick] (6.7,0.0) rectangle (9.6,1.9);
		\draw[rounded corners,blue2, ultra thick] (10.1,0.0) rectangle (13.0,1.9);
		\draw[rounded corners,red2, ultra thick] (-0.1,2.2) rectangle (2.8,4.1);
		\draw[rounded corners,red2, ultra thick] (3.3,2.2) rectangle (6.2,4.1);
		\draw[rounded corners,blue2, ultra thick] (6.7,2.2) rectangle (9.6,4.1);
		\draw[rounded corners,blue2, ultra thick] (10.1,2.2) rectangle (13.0,4.1);
		\node[text width=0cm] at (7.0,0.3) {\rotatebox{0}{$\hat{q}$}};
		\node[text width=0cm] at (10.3,0.3) {\rotatebox{0}{$\hat{q}$}};
		\node[text width=0cm] at (2.2,0.3) {\rotatebox{0}{$\hat{f}$}};
		\node[text width=0cm] at (5.6,0.3) {\rotatebox{0}{$\hat{f}$}};
		\node[text width=0cm] at (7.0,2.5) {\rotatebox{0}{$\hat{q}$}};
		\node[text width=0cm] at (10.3,2.5) {\rotatebox{0}{$\hat{q}$}};
		\node[text width=0cm] at (2.2,2.5) {\rotatebox{0}{$\hat{f}$}};
		\node[text width=0cm] at (5.6,2.5) {\rotatebox{0}{$\hat{f}$}};
		\draw[->,black, thick] (0.30,0.20) -- (0.70,0.20);   
		\draw[->,black, thick] (0.30,0.20) -- (0.30,0.60);   
		\node[text width=0cm] at (0.40,0.70) {$y$};
		\node[text width=0cm] at (0.80,0.15) {$x$};
		\scriptsize
		\draw[rounded corners, red2, fill=white, thick] (-0.3,2.55) rectangle (0.1,3.75);
		\node[text width=0cm] at (-0.2,3.15) {\rotatebox{90}{$\Lambda = 0^\circ$}};
		\draw[rounded corners, red2, fill=white, thick] (-0.3,0.35) rectangle (0.1,1.55);
		\node[text width=0cm] at (-0.2,0.95) {\rotatebox{90}{$\Lambda = 45^\circ$}};
		\draw[rounded corners, red2, fill=white, thick] (0.8,3.9) rectangle (1.95,4.3);
		\node[align=left] at (1.4,4.1) {\rotatebox{0}{$\alpha = 16^\circ$}};
		\draw[rounded corners, red2, fill=white, thick] (4.2,3.9) rectangle (5.35,4.3);
		\node[align=left] at (4.8,4.1) {\rotatebox{0}{$\alpha = 30^\circ$}};
		\draw[rounded corners, blue2, fill=white, thick] (7.6,3.9) rectangle (8.75,4.3);
		\node[align=left] at (8.2,4.1) {\rotatebox{0}{$\alpha = 16^\circ$}};
		\draw[rounded corners, blue2, fill=white, thick] (11.0,3.9) rectangle (12.15,4.3);
		\node[align=left] at (11.6,4.1) {\rotatebox{0}{$\alpha = 30^\circ$}};
		\end{tikzpicture} \\ 
		\centering
		\centering
		\begin{tikzpicture}
		\node[anchor=south west,inner sep=0] (image) at (0,0) {\includegraphics[width=1\textwidth]{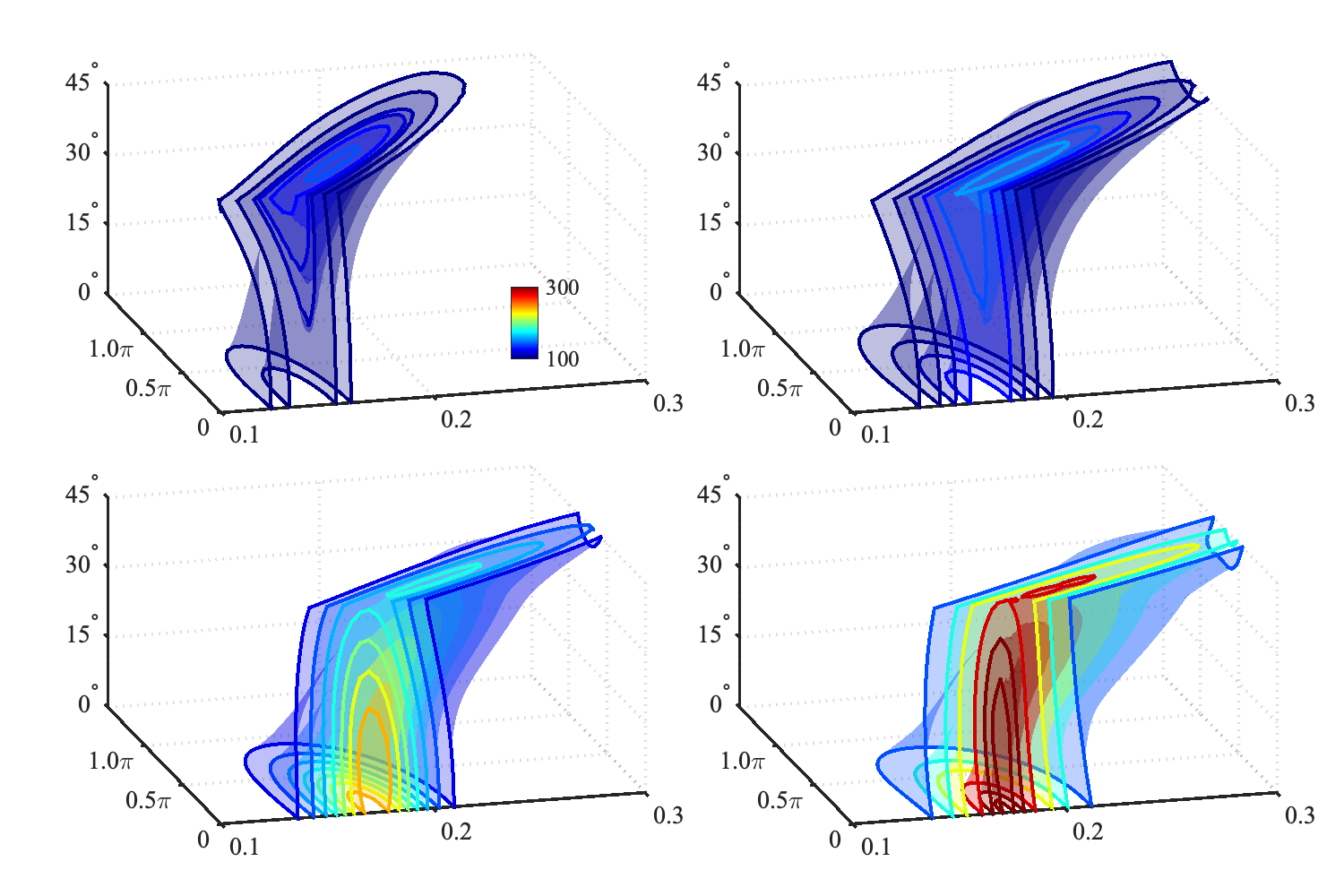}};
		\node[text width=0cm] at (0.3,7.0) {\rotatebox{90}{$\Lambda$}};
		\node[text width=0cm] at (0.3,3.0) {\rotatebox{90}{$\Lambda$}};
		\node[text width=0cm] at (0.4,1.0) {\rotatebox{0}{$k_{z^\prime}$}};
		\node[text width=0cm] at (0.4,5.0) {\rotatebox{0}{$k_{z^\prime}$}};
		\node[text width=0cm] at (5.1,6.3) {\rotatebox{0}{$\sigma_1$}};
		\node[text width=0cm] at (4.0,0.2) {\rotatebox{0}{$St^\prime$}};
		\node[text width=0cm] at (10.5,0.2) {\rotatebox{0}{$St^\prime$}};
		\node[text width=0cm] at (1.3,7.9) {\rotatebox{0}{$\alpha = 16^\circ$}};
		\node[text width=0cm] at (7.7,7.9) {\rotatebox{0}{$\alpha = 20^\circ$}};
		\node[text width=0cm] at (1.3,3.8) {\rotatebox{0}{$\alpha = 26^\circ$}};
		\node[text width=0cm] at (7.7,3.8) {\rotatebox{0}{$\alpha = 30^\circ$}};
		\end{tikzpicture} \\ \vspace{0mm}
		\caption{Forcing (in red boxes) and response (in blue boxes) contours for the $|u_x| / \|u_x\|_\infty \in [0.1,1]$ in blue-green-red scale at the largest $\sigma_1$ with $k_{z^\prime} = 0$ at $\Lambda = 0^\circ$ and $45^\circ$. Forcing modes extend in the wake and response modes become closer to the airfoil in swept wings. Isosurfaces of dominant resolvent gain $\sigma_1$ in the $\Lambda$--$St^\prime$--$k_{z^\prime}$ space for $\alpha = 16^\circ$ to $30^\circ$. } 
		\label{fig:resl_gainsIsosurf}
	\end{figure}
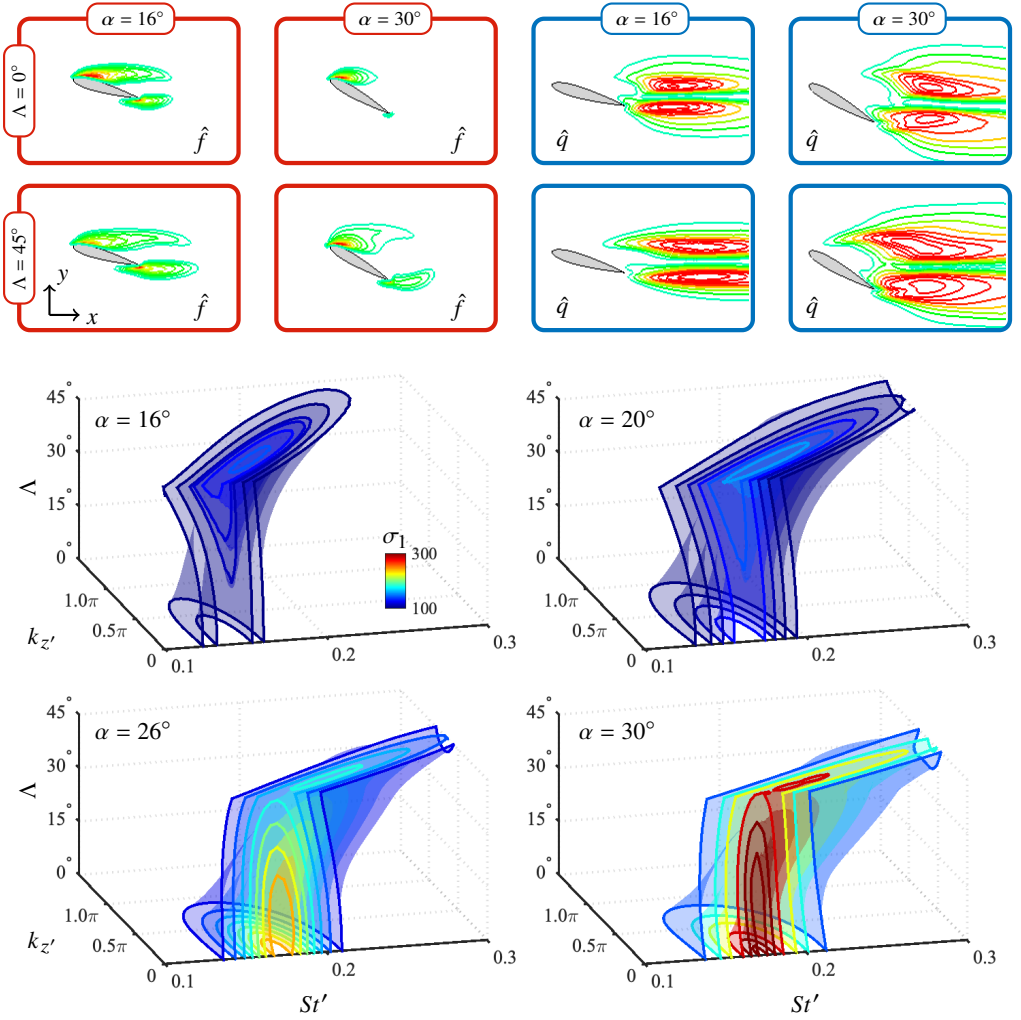

	To identify the existence of regions susceptible to the growth of perturbations in the flows over swept wings, we study the effect of sweep using resolvent analysis. Details on the stability of the linear operators and the usage of time-discounting are provided in the appendix \ref{sec:eigen}. We characterize the present flows through the dominant singular value $\sigma_1$ of the resolvent operator and its corresponding singular vectors $\hat{q}_{k_{z^\prime},\omega}$ and $\hat{f}_{k_{z^\prime},\omega}$. The influence of sweep on the vortex dynamics in the wake is analyzed through the forcing and response modes in figure \ref{fig:resl_gainsIsosurf}. The shown modal structures highlight the regions of the flow field which are more sensitive and responsive to the growth of perturbations. 
	
	Forcing modes are more localized than response modes, which are supported in the shear-dominated region of the flow, where perturbations can be highly amplified. On the other hand, the response modes are seen in the wake. For swept wings, the response modes are deformed spatially towards the airfoil surface and, for $\alpha = 30^\circ$, we notice the emergence of a characteristic responsive region near the airfoil leading edge. Such response structures  appear over swept wings only at high angles of attack, as seen in figure \ref{fig:resl_gainsIsosurf} (top).  We observe the contours of the magnitude of modal streamwise velocity component $| \hat{u}_x |$ to reveal the regions of the flow that have more responsive to introduced perturbations. We visualize similar contours for the forcing counterpart and see that the modes extend slightly into the wake, over the laminar separation bubble. This behavior reveals that the flow over swept wings can amplify optimal disturbances closer to the airfoil suction side, which can be used to alter the formation of the laminar separation bubble.
	
	Furthermore, the present resolvent analysis predicts the formation of oblique vortex shedding, as observed in \cite{Mittal:JFM21} and \cite{Zhang:JFM20b}, even though the present study is performed on spanwise periodic wings. Previous works have shown that oblique coherent structures become spatially periodic for large aspect ratio wings, making spanwise periodic analysis valid to study such three-dimensional structures. Through resolvent analysis, we can explain how oblique coherent structures are advected by the flow stream using the spatio-temporal frequencies of the optimal response modes.
	
	\begin{figure}
		\centering
		\begin{tikzpicture}
		\node[anchor=south west,inner sep=0] (image) at (0,0) {\includegraphics[trim=0mm 0mm 0mm 0mm, clip,width=1\textwidth]{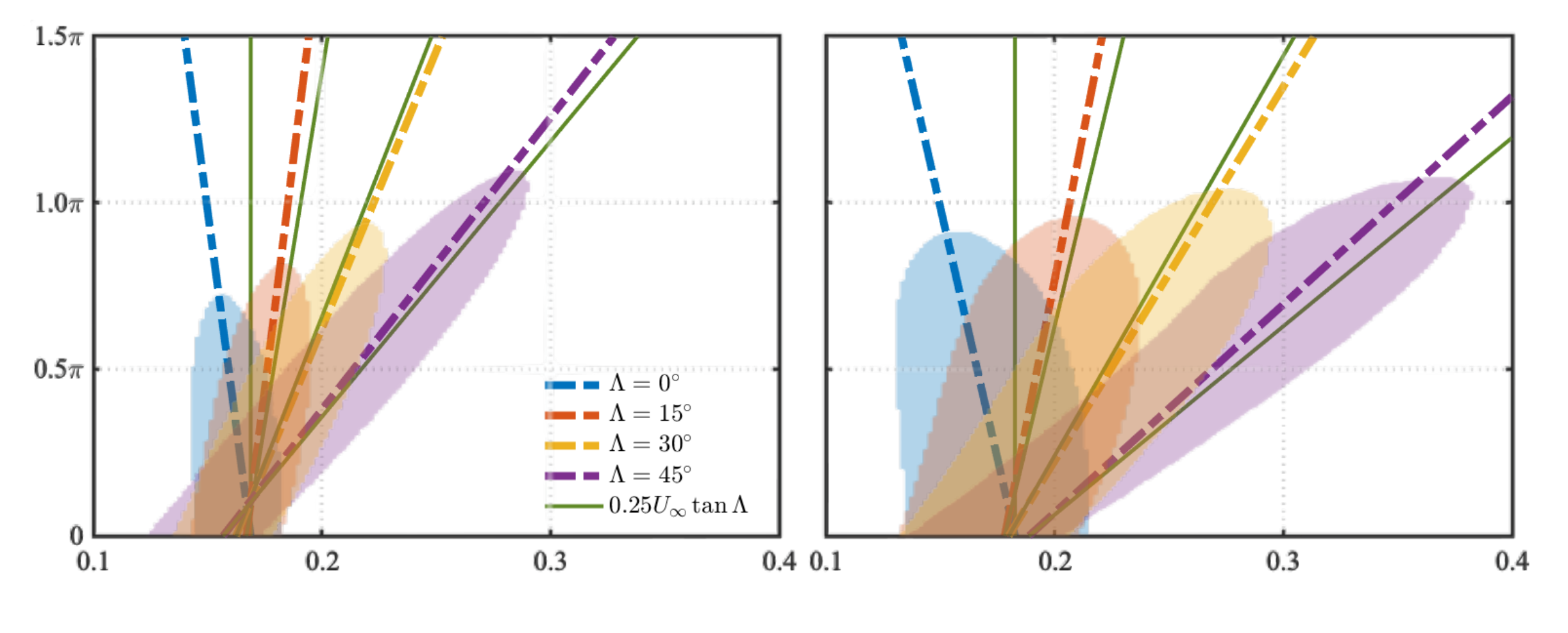}};
		\node[text width=0cm] at (0.0,3.0) {\rotatebox{90}{$k_{z^\prime}$}};
		\node[text width=0cm] at (3.8,0.2) {\rotatebox{0}{$St^\prime$}};
		\node[text width=0cm] at (10.1,0.2) {\rotatebox{0}{$St^\prime$}};
		\node[text width=0cm] at (5.6,4.7) {\rotatebox{0}{$\alpha = 20^\circ$}};
		\node[text width=0cm] at (11.8,4.7) {\rotatebox{0}{$\alpha = 30^\circ$}};
		\end{tikzpicture} \vspace{-7mm}
		\caption{The leading resolvent gain contours at $\alpha = 20^\circ$ and $30^\circ$ for $0^\circ \le \Lambda \le 45^\circ$. The dash-dotted slopes represent spanwise convection speeds. Green line exhibits the convection speed prediction with $0.25 U_\infty \tan \Lambda$.}
		\label{fig:resl_modalConvection}
	\end{figure}
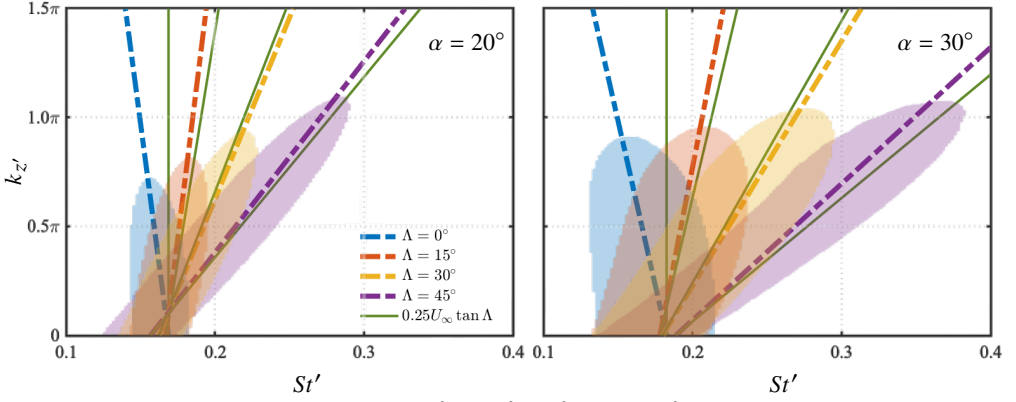
	
	The frequency at maximum $\sigma_1$ for each spanwise wavenumber is a function of the sweep angle and is characterized by the convection speed of the optimal oblique modes. We compute this phase speed as $c = \text{d} \omega / \text{d} k_{z^\prime}$, the slope of the optimal response frequencies for each spanwise wavenumber. This value is a function of $\Lambda$ and $k_{z^\prime}$ \citep{Paladini:PRF19,Plante:JFM21,HeTimme:JFM21}. For flow regimes studied in the present work, $f \approx 0.25 U_\infty \tan\Lambda$ gives a reasonable prediction for the frequency of the maximum $\sigma_1$ for each spanwise wavenumber and sweep angle for all angles of incidence, as shown in figure \ref{fig:resl_modalConvection}. This function can be used to predict the optimal forcing and response modes for laminar separated flows over swept wings. Additionally, the mode shapes of the optimal forcing and response are similar for low $k_{z^\prime}$. The present results reveal the optimal actuation location and response as well as the spatio-temporal behavior of the flow perturbations over laminar separated flows on swept wings. 
	
	In general, oblique modes are the most amplified optimal disturbances for all swept wings. The effect of sweep on $\sigma_1$, however, depends on the angle of attack, as shown in figure \ref{fig:resl_gainsIsosurf} and summarized in table \ref{tab:resl_max_norm}. For $\alpha \le 20^\circ$, swept wings have higher amplification than unswept wings. This is a distinct behavior when compared to $\alpha \ge 26^\circ$, in which swept wings have lower resolvent norm than unswept wings. This behavior suggests that it is more challenging to perturb highly swept wings at high angles of attack.
	
	\renewcommand{\arraystretch}{1.5}
	\begin{table}
		\flushleft
		\begin{tabular}{p{0.21in}p{0.4in}p{0.4in}p{0.3in}p{0.04in}p{0.4in}p{0.4in}p{0.3in}}
			\multicolumn{1}{l}{} & 
			\multicolumn{3}{l}{\hspace{11mm}$\alpha = 16^\circ$} && \multicolumn{3}{l}{\hspace{11mm}$\alpha = 20^\circ$} \\
			\multicolumn{1}{l}{$\Lambda$} & \multicolumn{1}{l}{\hspace{0mm}${\rm max}(\sigma_1)$} & \multicolumn{1}{l}{\hspace{2mm}$St^\prime$} & \multicolumn{1}{l}{\hspace{2mm}$k_{z^\prime}$} && \multicolumn{1}{l}{\hspace{0mm}${\rm max}(\sigma_1)$} & \multicolumn{1}{l}{\hspace{2mm}$St^\prime$} & \multicolumn{1}{l}{\hspace{2mm}$k_{z^\prime}$} \\ 
			\cline{2-4} \cline{6-8} 
			$0^\circ$ & $107.7$ & $0.143$ & $0.000 \pi$ && $137.0$ & $0.165$ & $0.000 \pi$ \\
			$15^\circ$ & $110.7$ & $0.146$ & $0.161 \pi$ && $140.6$ & $0.169$ & $0.171 \pi$ \\
			$30^\circ$ & $121.3$ & $0.154$ & $0.281 \pi$ && $150.3$ & $0.175$ & $0.221 \pi$ \\
			$45^\circ$ & $145.3$ & $0.170$ & $0.392 \pi$ && $166.5$ & $0.190$ & $0.302 \pi$ \\
			\hline
			\hline
			\multicolumn{1}{l}{$\alpha$} & 
			\multicolumn{3}{l}{\hspace{11mm}$\alpha = 26^\circ$} && \multicolumn{3}{l}{\hspace{11mm}$\alpha = 30^\circ$} \\
			\multicolumn{1}{l}{$\Lambda$} & \multicolumn{1}{l}{\hspace{0mm}${\rm max}(\sigma_1)$} & \multicolumn{1}{l}{\hspace{2mm}$St^\prime$} & \multicolumn{1}{l}{\hspace{2mm}$k_{z^\prime}$} && \multicolumn{1}{l}{\hspace{0mm}${\rm max}(\sigma_1)$} & \multicolumn{1}{l}{\hspace{2mm}$St^\prime$} & \multicolumn{1}{l}{\hspace{2mm}$k_{z^\prime}$} \\ 
			\cline{2-4} \cline{6-8} 
			$0^\circ$ & $250.1$ & $0.173$ & $0.000 \pi$ && $368.5$ & $0.174$ & $0.000 \pi$ \\
			$15^\circ$ & $257.2$ & $0.178$ & $0.131 \pi$ && $375.9$ & $0.175$ & $0.090 \pi$ \\
			$30^\circ$ & $228.2$ & $0.189$ & $0.191 \pi$ && $363.8$ & $0.182$ & $0.101 \pi$ \\
			$45^\circ$ & $186.6$ & $0.207$ & $0.211 \pi$ && $299.4$ & $0.195$ & $0.111 \pi$
		\end{tabular}
		\flushright
		\begin{tikzpicture}
		\node[overlay,remember picture,anchor=south west,inner sep=0] (image) at (-5,0.2) {\includegraphics[trim=0mm 15mm 0mm 0mm,clip,width=0.4\textwidth]{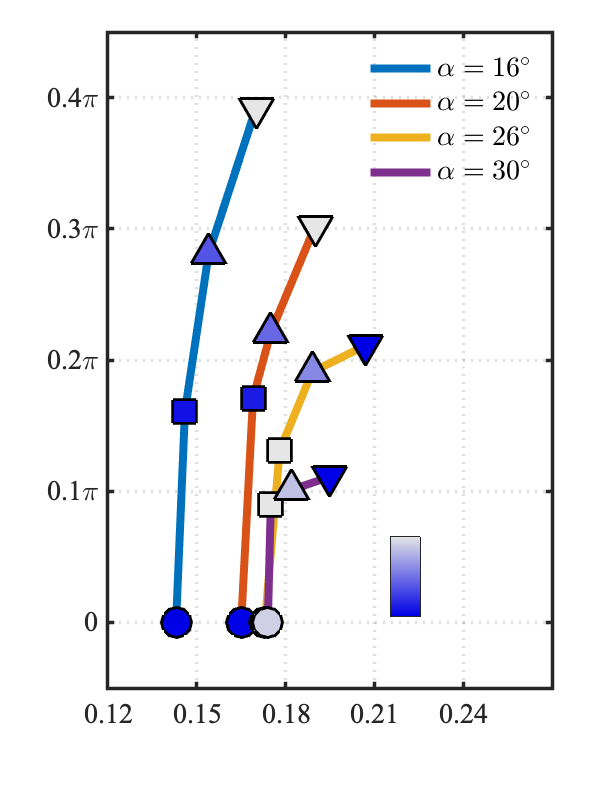}};
		\node[overlay,remember picture,text width=5cm] at (-2.55,3.6) {\rotatebox{90}{$k_{z^\prime}$}};
		\node[overlay,remember picture,text width=5cm] at (0.2,0.2) {$St^\prime$};
		\tiny
		\node[overlay,remember picture,text width=5cm] at (1.4,2.1) {$\sigma_{1,{\rm max}}$};
		\node[overlay,remember picture,text width=5cm] at (1.4,1.4) {$\sigma_{1,{\rm min}}$};
		\tiny
		\node[overlay,remember picture,text width=5cm] at (1.1,4.11) {\large $\circ$};
		\node[overlay,remember picture,text width=5cm] at (1.4,4.13) {$\Lambda = 0^\circ$};
		\node[overlay,remember picture,text width=5cm] at (1.1,3.82) {\tiny $\square$};
		\node[overlay,remember picture,text width=5cm] at (1.4,3.83) {$\Lambda = 15^\circ$};
		\node[overlay,remember picture,text width=5cm] at (1.09,3.51) {\tiny $\triangle$};
		\node[overlay,remember picture,text width=5cm] at (1.4,3.55) {$\Lambda = 30^\circ$};
		\node[overlay,remember picture,text width=5cm] at (1.1,3.25) {\scriptsize $\triangledown$};
		\node[overlay,remember picture,text width=5cm] at (1.4,3.25) {$\Lambda = 45^\circ$};
		\end{tikzpicture}
		\caption{The maximum leading resolvent gain ${\rm max}(\sigma_1)$ for each $\alpha,\Lambda$ pair. On the right, we plot ${\rm max}(\sigma_1)$ in $St^\prime$--$k_{z^\prime}$ space colored in blue scale with respect to the minimum and maximum $\sigma_1$ for each $\alpha$.}
		\label{tab:resl_max_norm}
	\end{table} 
	
	\begin{figure}
		\centering
		\begin{tikzpicture}
		\node[anchor=south west,inner sep=0] (image) at (2.2,-9) {\includegraphics[trim=9mm 2mm 2mm 0mm, clip,width=0.7\textwidth]{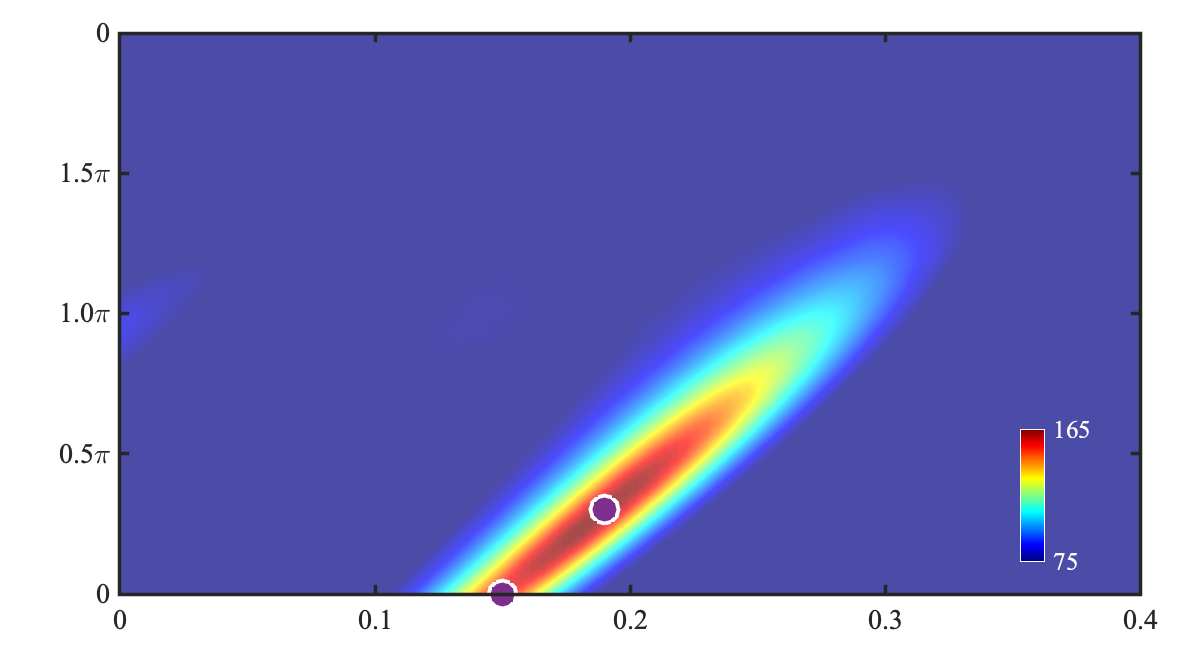}};
		\draw[rounded corners,red2, fill=white, ultra thick] (0.5,-5.6) rectangle (3.2,-1.6);
		\draw[rounded corners,blue2,fill=white, ultra thick] (3.25,-5.6) rectangle (6.05,-1.6);
		\draw[rounded corners,red2,fill=white, ultra thick] (8.0,-5.6) rectangle (10.7,-1.6);
		\draw[rounded corners,blue2,fill=white, ultra thick] (10.75,-5.6) rectangle (13.55,-1.6);
		\node[anchor=south west,inner sep=0] (image) at (0.5,-5.55) {\includegraphics[trim=0mm 2mm 25mm 0mm, clip,height=0.3\textwidth]{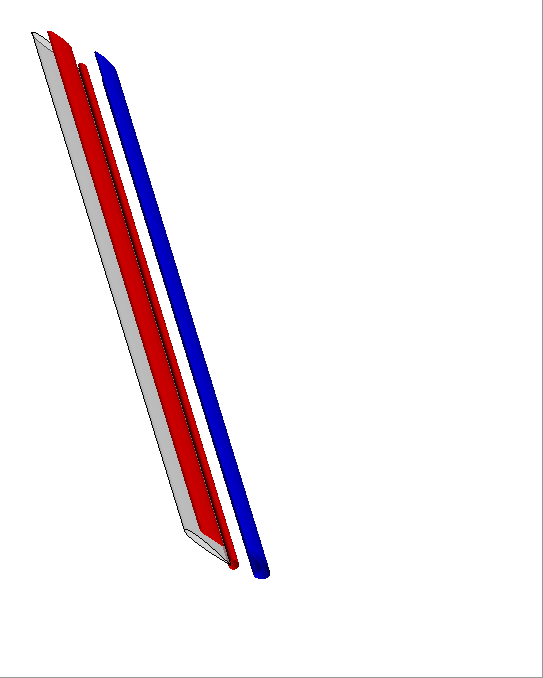}};
		\node[anchor=south west,inner sep=0] (image) at (3.2,-5.55) {\includegraphics[trim=0mm 2mm 25mm 0mm, clip,height=0.3\textwidth]{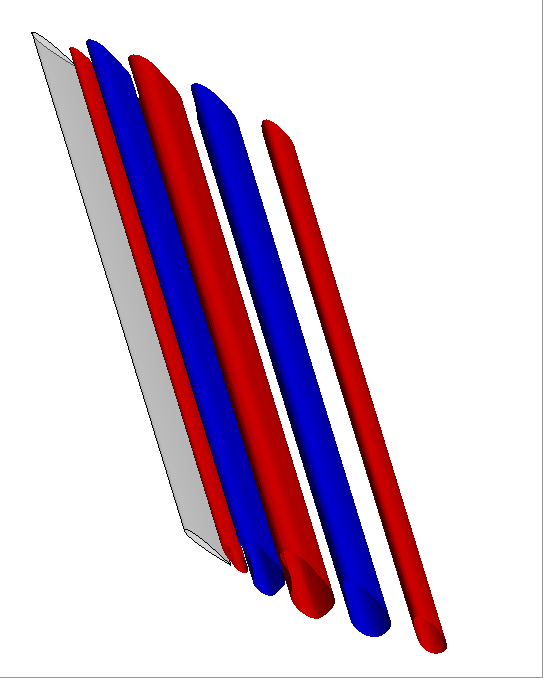}};
		\node[anchor=south west,inner sep=0] (image) at (8.0,-5.55) {\includegraphics[trim=0mm 2mm 25mm 0mm, clip,height=0.3\textwidth]{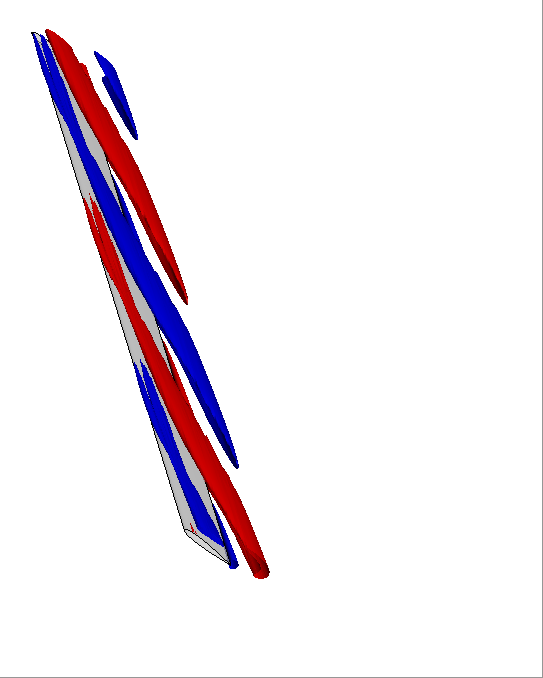}};
		\node[anchor=south west,inner sep=0] (image) at (10.7,-5.55) {\includegraphics[trim=0mm 2mm 25mm 0mm, clip,height=0.3\textwidth]{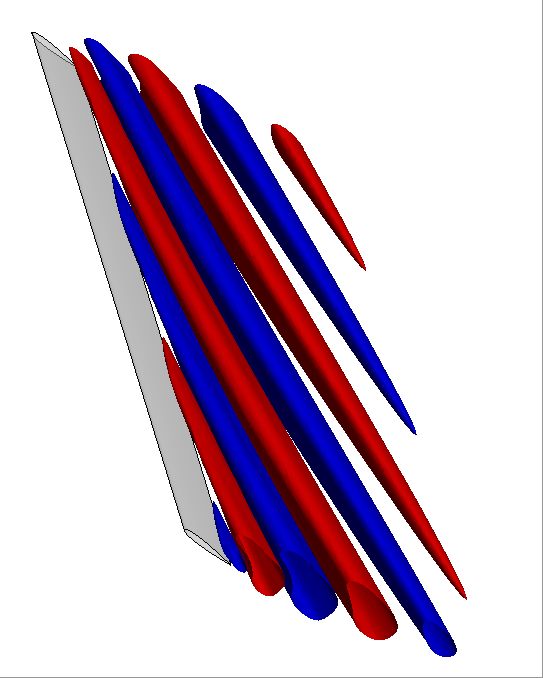}};
		\draw[rounded corners,red2, ultra thick] (0.5,-5.6) rectangle (3.2,-1.6);
		\draw[rounded corners,blue2,ultra thick] (3.25,-5.6) rectangle (6.05,-1.6);
		\draw[rounded corners,red2, ultra thick] (8.0,-5.6) rectangle (10.7,-1.6);
		\draw[rounded corners,blue2,ultra thick] (10.75,-5.6) rectangle (13.55,-1.6);
		\node[text width=0cm] at (6.9,-9.1) {\rotatebox{0}{$St^\prime$}};
		\node[text width=0cm] at (2.0,-6.7) {\rotatebox{90}{$k_{z^\prime}$}};
		\node[text width=0cm] at (10.3,-6.8) {\rotatebox{0}{\color{white}$\sigma_1$}};	
		\normalsize
		\draw[<->,white, ultra thick] (5.95,-8.5) .. controls (3.225,-7.0) .. (3.225,-5.525);
		\draw[<->,purple2, thick] (5.925,-8.475) .. controls (3.225,-7.0) .. (3.225,-5.55);
		\draw[<->,white, ultra thick] (7.05,-7.8) .. controls (10.725,-6.0) .. (10.725,-5.525);
		\draw[<->,purple2, thick] (7.075,-7.775) .. controls (10.725,-6.0) .. (10.725,-5.55);
		\node[text width=0cm] at (0.75,-5.2) {\rotatebox{0}{$\hat{f}$}};
		\node[text width=0cm] at (3.5,-5.2) {\rotatebox{0}{$\hat{q}$}};
		\node[text width=0cm] at (8.25,-5.2) {\rotatebox{0}{$\hat{f}$}};
		\node[text width=0cm] at (11.0,-5.2) {\rotatebox{0}{$\hat{q}$}};
		\end{tikzpicture} \vspace{-4mm}
		\caption{Forcing ($\hat{f}$, in red boxes) and response ($\hat{q}$, in blue boxes) modes isosurfaces with $y$--velocity components $\hat{u}_y / \| \hat{u}_y \|_\infty \in \pm 0.2$ in red-blue color scale for $\alpha = 20^\circ$ and $\Lambda = 45^\circ$. Wingspan length is $10\ L_c$. Forcing and response modes are associated  with the largest resolvent gain for each $k_{z^\prime}$ as shown in the $\sigma_1$ contours over $St^\prime$--$k_{z^\prime}$ plane.} 
		\label{fig:resl_modes_aoa20sw45}
	\end{figure}
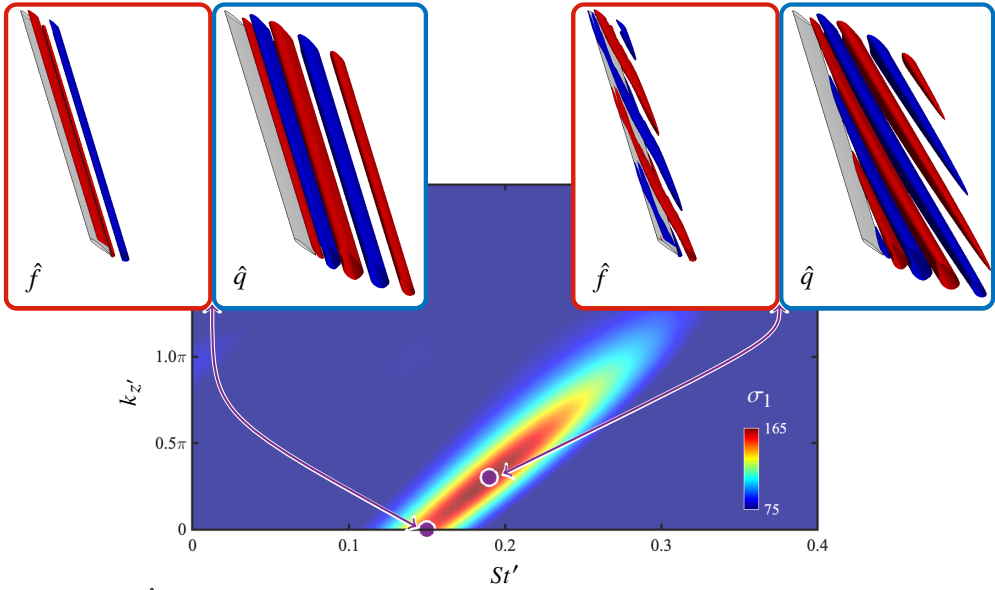

	The spatial and temporal frequency of the maximum resolvent gain ${\rm max}(\sigma_1)$ in the  $St^\prime$--$k_{z^\prime}$ space depends on the sweep angle, as shown in table \ref{tab:resl_max_norm}. For unswept wings, the largest resolvent gain $\sigma_1$ is found for the $2$D setting of $k_{z^\prime} = 0$ associated with the temporal frequency of the characteristic vortex shedding. However, both $k_{z^\prime}$ and $St^\prime$ of the optimal disturbances ${\rm max}(\sigma_1)$ increase with the sweep angle. Thus, the $3$D oblique modal structures are not only predicted by the present resolvent analysis but also found to be the most amplified flow mechanism in swept wings, as shown in figure \ref{fig:resl_modes_aoa20sw45}.
	
	Although all flows analyzed in this work are spanwise periodic, and oblique shedding is absent in the DNS, the large $\sigma_1$ in $k_{z^\prime} > 0$ modes suggest that the spanwise flow over swept wings supports the formation and shedding of $3$D oblique vortices in infinite wings. To analyze the spatial behavior of oblique modes, let us focus on the resolvent analysis at the angle of attack $\alpha = 20^\circ$ and sweep angle $\Lambda = 45^\circ$, as seen in figure \ref{fig:resl_modes_aoa20sw45}.  As noticed at $k_{z^\prime} = 0$, the 2D forcing and response modes are aligned with the wingspan, however, the maximum resolvent gain $\sigma_1$ in the $St^\prime$--$k_{z^\prime}$ for this flow is found at $k_{z^\prime} = 0.3\pi$ and $St = 0.19$, where modes are oblique  with respect to the wingspan. For this reason, the flow over swept wings has a higher propensity to develop oblique shedding when compared to the flow over unswept wings and such characteristic is revealed through resolvent analysis to be associated with the sweep angle.
	
	The flow mechanisms that are responsible for oblique shedding and the attenuation of the spanwise oscillations for swept wings were described as the growth of response modes towards the airfoil surface and the extension of forcing modes into the wake and over the laminar separation bubble. This phenomena also creates an overlapping region of the flow where both forcing and response modes are supported. This overlap of forcing and response structures is more prominent at the higher sweep angles, although it is also present in unswept wings. The overlap of forcing and response modes and their associated resolvent gain can both be relevant to characterize how the flow over swept wings at high incidence give rise to perturbations on the flow as we discuss such phenomena through the lens of wavemakers.
	
	\subsection{Wavemakers}
	\label{sec:wavemakers}
	
	Wavemakers have been described as regions of the flow field characterized by both high sensitivity and responsiveness to perturbation growth \citep{Giannetti:JFM07,Giannetti:JFM10,FosasdePando:JFM17}.  Such regions are optimal for the introduction of self-sustained instabilities, acting as the source of global instabilities of the system, and motivate the analysis of structural sensitivity of the modal forcing and response structures. 
	
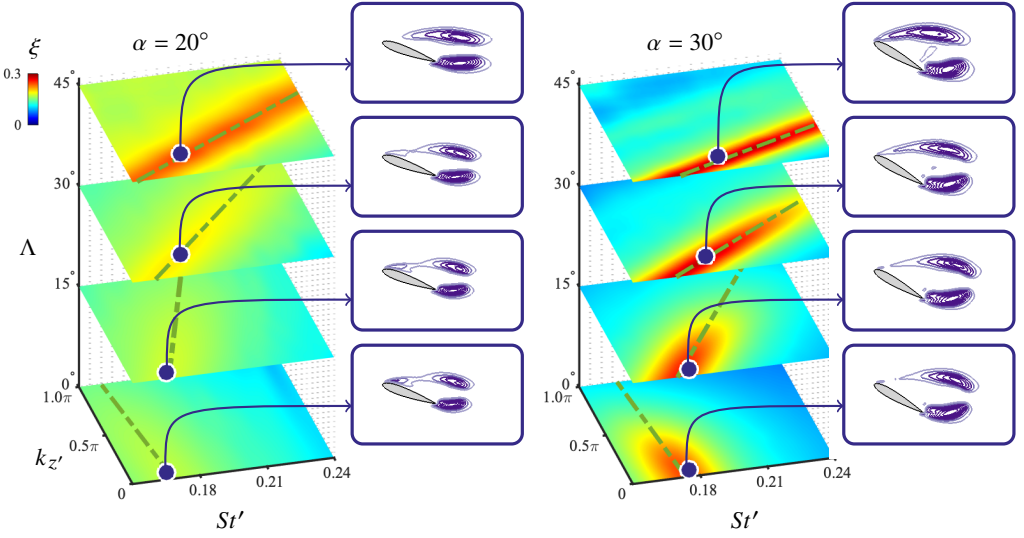
\begin{figure}
		\centering
		\begin{tikzpicture}
		\node[anchor=south west,inner sep=0] (image) at (-0.3,0.0) {\includegraphics[trim=0mm 0mm 0mm 0mm, clip,width=1\textwidth]{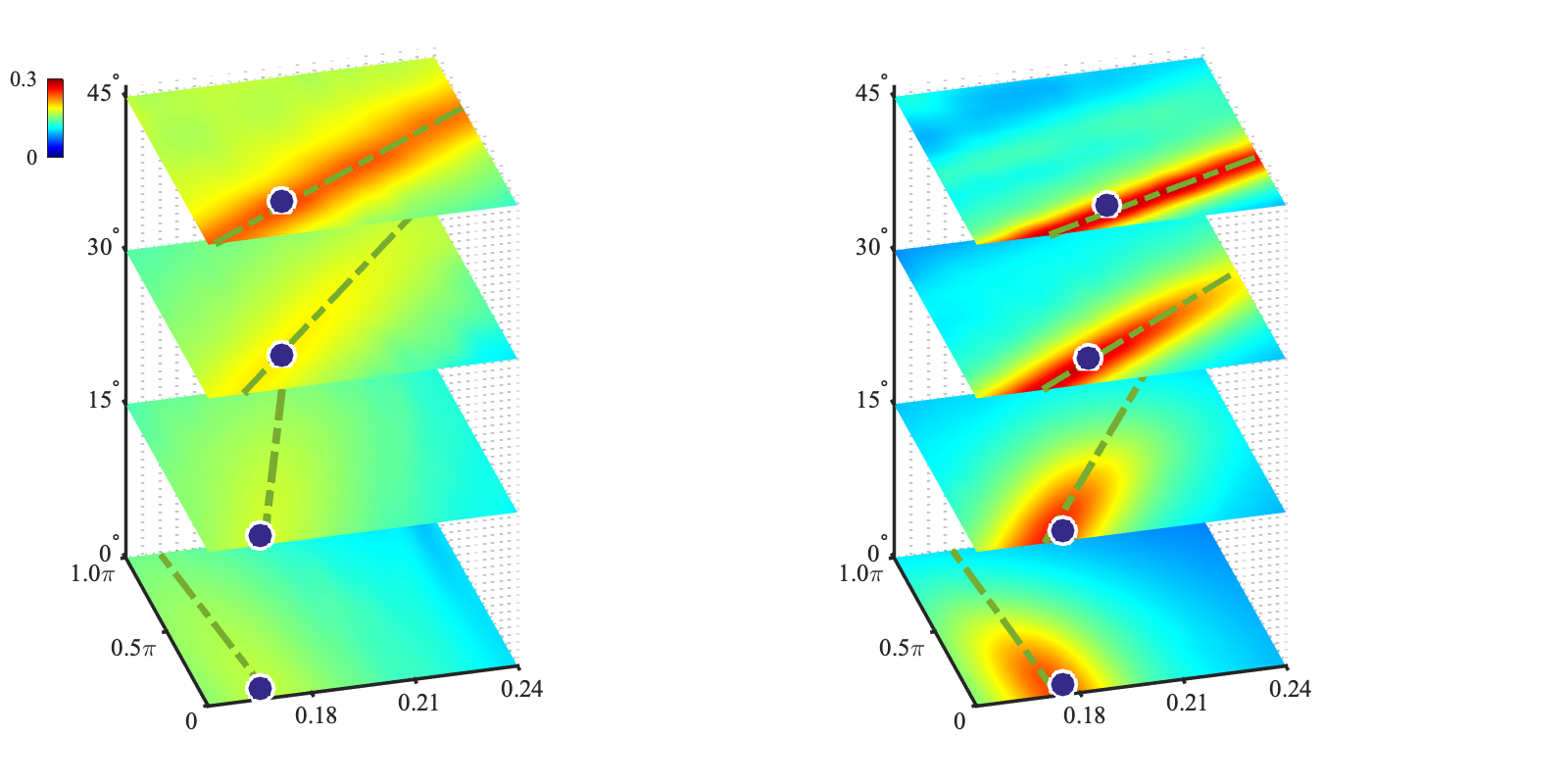}};
		\node[anchor=south west,inner sep=0] (image) at (4.2,5.5) {\includegraphics[trim=0mm 0mm 0mm 0mm, clip,width=0.18\textwidth]{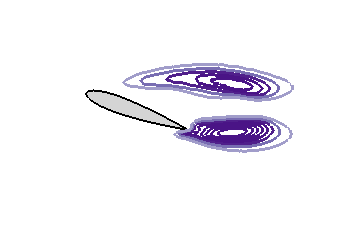}};
		\node[anchor=south west,inner sep=0] (image) at (4.2,4.0) {\includegraphics[trim=0mm 0mm 0mm 0mm, clip,width=0.18\textwidth]{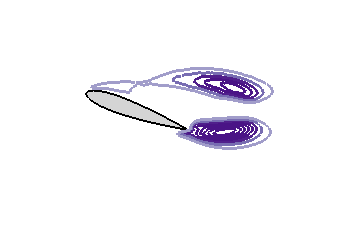}};
		\node[anchor=south west,inner sep=0] (image) at (4.2,2.5) {\includegraphics[trim=0mm 0mm 0mm 0mm, clip,width=0.18\textwidth]{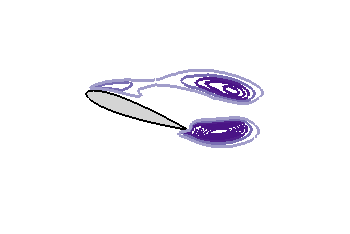}};
		\node[anchor=south west,inner sep=0] (image) at (4.2,1.0) {\includegraphics[trim=0mm 0mm 0mm 0mm, clip,width=0.18\textwidth]{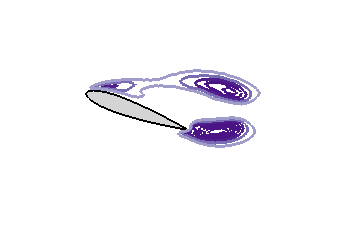}};
		\node[anchor=south west,inner sep=0] (image) at (10.7,5.5) {\includegraphics[trim=0mm 0mm 0mm 0mm, clip,width=0.18\textwidth]{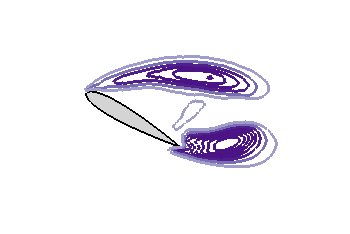}};
		\node[anchor=south west,inner sep=0] (image) at (10.7,4.0) {\includegraphics[trim=0mm 0mm 0mm 0mm, clip,width=0.18\textwidth]{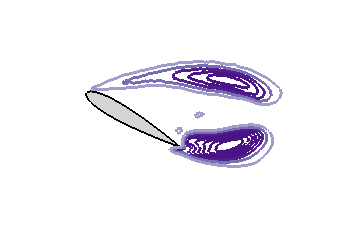}};
		\node[anchor=south west,inner sep=0] (image) at (10.7,2.5) {\includegraphics[trim=0mm 0mm 0mm 0mm, clip,width=0.18\textwidth]{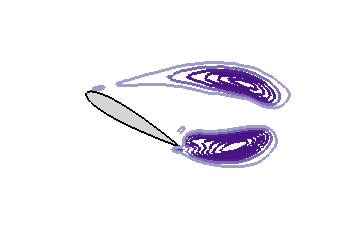}};
		\node[anchor=south west,inner sep=0] (image) at (10.7,1.0) {\includegraphics[trim=0mm 0mm 0mm 0mm, clip,width=0.18\textwidth]{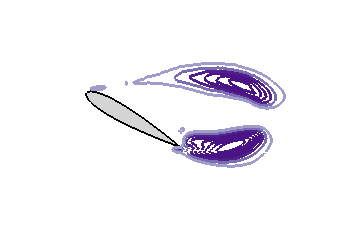}};
		\node[text width=0cm] at (0.13,6.4) {\rotatebox{0}{$\xi$}};
		\node[text width=0cm] at (0.0,3.8) {\rotatebox{0}{$\Lambda$}};
		\node[text width=0cm] at (0.2,1.0) {\rotatebox{0}{$k_{z^\prime}$}};
		\node[text width=0cm] at (2.6,0.2) {\rotatebox{0}{$St^\prime$}};
		\node[text width=0cm] at (9.4,0.2) {\rotatebox{0}{$St^\prime$}};
		\node[text width=0cm] at (1.5,6.5) {\rotatebox{0}{$\alpha = 20^\circ$}};
		\node[text width=0cm] at (8.3,6.5) {\rotatebox{0}{$\alpha = 30^\circ$}};
		\large
		\draw[->,jetI, thick] (2.13,5.02) .. controls (2.13,6.2) .. (4.4,6.2);
		\draw[rounded corners, jetI, very thick] (4.4,5.7) rectangle (6.65,7.0);
		\draw[->,jetI, thick] (2.12,3.62) .. controls (2.12,4.6) .. (4.4,4.6);
		\draw[rounded corners, jetI, very thick] (4.4,4.2) rectangle (6.65,5.5);
		\draw[->,jetI, thick] (1.95,2.22) .. controls (1.95,3.1) .. (4.4,3.1);
		\draw[rounded corners, jetI, very thick] (4.4,2.7) rectangle (6.65,4.0);
		\draw[->,jetI, thick] (1.93,0.88) .. controls (1.93,1.7) .. (4.4,1.7);
		\draw[rounded corners, jetI, very thick] (4.4,1.2) rectangle (6.65,2.5);
		\draw[->,jetI, thick] (9.24,5.02) .. controls (9.24,6.2) .. (10.9,6.2);
		\draw[rounded corners, jetI, very thick] (10.9,5.7) rectangle (13.15,7.0);
		\draw[->,jetI, thick] (9.08,3.62) .. controls (9.08,4.6) .. (10.9,4.6);
		\draw[rounded corners, jetI, very thick] (10.9,4.2) rectangle (13.15,5.5);
		\draw[->,jetI, thick] (8.84,2.22) .. controls (8.84,3.1) .. (10.9,3.1);
		\draw[rounded corners, jetI, very thick] (10.9,2.7) rectangle (13.15,4.0);
		\draw[->,jetI, thick] (8.82,0.88) .. controls (8.82,1.7) .. (10.9,1.7);
		\draw[rounded corners, jetI, very thick] (10.9,1.2) rectangle (13.15,2.5);
		\end{tikzpicture} \vspace{-5mm}
		\caption{Contours of $\xi = (\sigma_1 / {\rm max}(\sigma_1)) \langle \hat{q}_i, \hat{f}_i \rangle$, the inner product between forcing and response modes, scaled by the ratio of $\sigma_1$ and the  maximum $\sigma_1$ for each $\alpha$. Green line shows the convection speed $c = \text{d} \omega / \text{d} k_{z^\prime}$ for the optimal wavemakers. Spatial modes shown by the Hadamard product of $\hat{q}_i$ and $\hat{f}_i$, in magnitude, normalized by their maximum value, and colored in purple scale.} 
		\label{fig:resl_wavemaker}
	\end{figure}

	In global stability analysis, wavemakers are generally derived with direct and adjoint modes. Here, we quantify the strength of the wavemakers with the inner product of the forcing and response modes $\langle \hat{q}_{k_{z^\prime},\omega}, \hat{f}_{k_{z^\prime},\omega} \rangle$ and visualize the corresponding wavemaker modes with their Hadamard product. Wavemaker modes exhibit a higher magnitude downstream the airfoil, as shown in figure \ref{fig:resl_wavemaker}, in the region where $3$D flow develops for $\alpha = 30^\circ$. Thus, the emergence of strong wavemakers near the leading edge and over the separation bubble highlights the presence of self-sustained oscillations in the flow field. As the sweep angle tends to empower forcing and response modes overlap, wavemakers tend to be spatially wider in swept wings.
	
	We evaluate the strength of wavemakers in the $\Lambda$--$St^\prime$--$k_{z^\prime}$ space with the inner product of pseudomodes $\langle \hat{q}_{k_{z^\prime},\omega}, \hat{f}_{k_{z^\prime},\omega} \rangle$ and their associated resolvent gain $\sigma_1$. In this way, we avoid accounting for the wavemakers where $\sigma_1$ is too small to amplify perturbations. Hence, to understand which combination of temporal frequencies, spanwise wavenumber, and sweep angle is most likely to generate wavemakers, we consider the coefficient $\xi = (\sigma / {\rm max} (\sigma_1)) \langle \hat{q}_i, \hat{f}_i \rangle$, where $ {\rm max} (\sigma_1)$ is evaluated for each angle of attack over the $St^\prime$--$k_{z^\prime}$--$\Lambda$ space, with contours shown in figure \ref{fig:resl_wavemaker} for $\alpha = 20^\circ$ and $30^\circ$. 
	
	The resolvent modes with higher values of $\xi$ are observed for swept wings at higher angles of attack.  The wavemaker modes appear where vortex shedding develops in the wake. Hence a higher $\xi$ suggests perturbations are introduced with higher gain to be amplified in this region for flows over wings at high incidence. Furthermore, those disturbances feed the flow with self-generated disturbances that maintain three-dimensionality of the wake, as observed for instance at $\alpha = 30^\circ$. In unswept wings, the $St^\prime$--$k_{z^\prime}$ frequencies with strong wavemaker modes is found for $2$D wavemakers.
	
	For swept wings, the modes with the highest $\xi$ coefficient for each angles of attack are located at $\Lambda = 45^\circ$ and nonzero $k_{z^\prime}$, hence being associated with oblique modes. This finding is in agreement with the previous observations on the overlap of forcing and response modes, in figure \ref{fig:resl_gainsIsosurf}.  Hence, even if the amplification gain $\sigma_1$ is reduced for swept wings at $\alpha = 30^\circ$, the overlap of forcing and response modes is stronger, which introduces wavemakers over swept wings that are stronger than wavemakers for unswept wings, which explains the three-dimensionality observed in these flow fields.
		
	This finding, however, is in contrast with the DNS results that show an attenuation of spanwise oscillations with the sweep angle. To understand why such alleviation occurs, we must observe that optimal responses and optimal wavemakers also have an associated wave speed, characterized by their spatial and temporal frequencies, that is associated with the transport of disturbances over the periodic direction.
	
	%
	%
	%
	%
	%
	%
	%
	%
	%
	\renewcommand{\arraystretch}{1.5}
	\begin{table}
		\centering
		\begin{tabular}{p{0.95in}p{0.001in}p{0.39in}p{0.39in}p{0.39in}p{0.39in}p{0.001in}p{0.39in}p{0.39in}p{0.39in}p{0.39in}}
			\multicolumn{1}{r}{} &&
			\multicolumn{4}{c}{\hspace{0mm}$\alpha = 26^\circ$} &&
			\multicolumn{4}{c}{\hspace{0mm}$\alpha = 30^\circ$} \\
			&&
			\multicolumn{1}{c}{\hspace{0mm}$\Lambda = 0^\circ$} &
			\multicolumn{1}{c}{\hspace{0mm}$15^\circ$} &
			\multicolumn{1}{c}{\hspace{0mm}$30^\circ$} &
			\multicolumn{1}{c}{\hspace{0mm}$45^\circ$} &&
			\multicolumn{1}{c}{\hspace{0mm}$0^\circ$} &
			\multicolumn{1}{c}{\hspace{0mm}$15^\circ$} &
			\multicolumn{1}{c}{\hspace{0mm}$30^\circ$} &
			\multicolumn{1}{c}{\hspace{0mm}$45^\circ$} \\
			\cline{3-6}\cline{8-11}
			Opt. response && $-0.021$ & $0.022$ & $0.089$ & $0.253$ && $-0.042$ & $0.039$ & $0.150$ & $0.406$ \\
			Opt. wavemaker && $0.117$ & $0.185$ & $0.316$ & $0.595$ && $-0.008$ & $0.073$ & $0.204$ & $0.537$  \\
		\end{tabular}
		\caption{Convective speed $c = \text{d} \omega / \text{d} k_{z^\prime}$ for the optimal response and the optimal wavemakers for the $3$D flows at $\alpha = 26^\circ$ and $30^\circ$ and sweep angles $0^\circ \le \Lambda \le 45^\circ$.}
		\label{tab:wavemakers}
	\end{table}

	We note that an optimal wavemaker speed being faster than the optimal response wave speed leads to an attenuation on three-dimensionality. Wavemakers yield self-sustained instabilities in swept wings with an associated wavemaker phase speed $c_w = \text{d}\omega / \text{d} k_{z^\prime}$, which we characterize by the slope of the slash-dotted green lines in figure \ref{fig:resl_wavemaker}.  As observed in table \ref{tab:wavemakers}, when $c_w$ is large for high sweep angles and the optimal response $c$ is small, the reduction of spanwise oscillations is seen in the flowfield. In such cases, a misalignment appears between optimal responses and wavemakers which can not support spanwise oscillations. For this reason, wavemakers cannot sustain three-dimensional disturbances over swept wings.
		
	Finally, even for $\alpha = 20^\circ$, in which the flow field is $2$D, resolvent analysis reveals the presence of wavemakers. Those are associated with the sustained unsteady $2$D vortex shedding. To sustain three-dimensionality, wavemakers must introduce sufficiently strong three-dimensional structures to the flow with high amplification gain. For swept wings at high incidence, even though optimal wavemakers have a high gain, they are advected faster than the optimal responses, which reduces the flow three-dimensionality.
	
	\section{Conclusions}
	\label{sec:conclusions}
	We reported on the wake dynamics under the influence of sweep for laminar flows over two-dimensional wings through the use of direct numerical simulations and resolvent analysis. The study focused on the onset of $3$D wake structures at high incidence and the reduction of spanwise oscillations at high sweep.  DNS revealed the influence of sweep in terms of attenuating spanwise fluctuations over the wing and giving rise to three-dimensional wakes in agreement with the literature on finite wings. Although the wake dynamics exhibit larger differences between swept and unswept wings, a sweep-angle based scaling can be used to collapse aerodynamic characteristics when we consider streamwise and spanwise flows to be independent.
	
	As some differences in pressure, lift, and drag for the lower angle of attack settings are perceived at the higher angles of sweep and attack, we resort to the force element theory and identify the vortical structures with spanwise periodicity formed closer to the wing within the laminar separation bubble. Such elements are observed to increase in size and shape with an increase in angles of sweep and attack, a behavior associated with the deviations of scaled force and pressure coefficients for massively separated flows. This finding revealed force elements that impose additional forces over the wing  and showed that spanwise and streamwise flow components cannot be independently analyzed for massively separated flows over swept wings.
	
	Through resolvent analysis, we showed how sweep angle induces a convection speed to the optimal resolvent modes and provide a linear model to predict the optimal forcing and response spatio-temporal frequencies for laminar flows over swept wings. The forcing and response spatial mode pairs are also affected by the sweep as  well as wavemakers that sustain and promote unsteadiness in the wake. We revealed that a misalignment between the optimal response convection speed and the wavemaker speed leads to the reduction in spanwise oscillations for higher sweep angle. Additionally, we observed that resolvent modes with large amplification gain on swept wings represent the oblique vortex shedding, as observed for laminar flows over high aspect ratio wings in the literature. The present results reveal the fundamental influence of the sweep angle on the airfoil wake dynamics and support future studies on the control of wake oscillations on swept wings at higher angles of attack. 
	
	\appendix
	\section{Dominant eigenvalues of the linearized operators}
	\label{sec:eigen}
	
\begin{figure}
		\centering
		\begin{tikzpicture}
		\node[anchor=south west,inner sep=0] (image) at (0,0) {\includegraphics[width=1\textwidth]{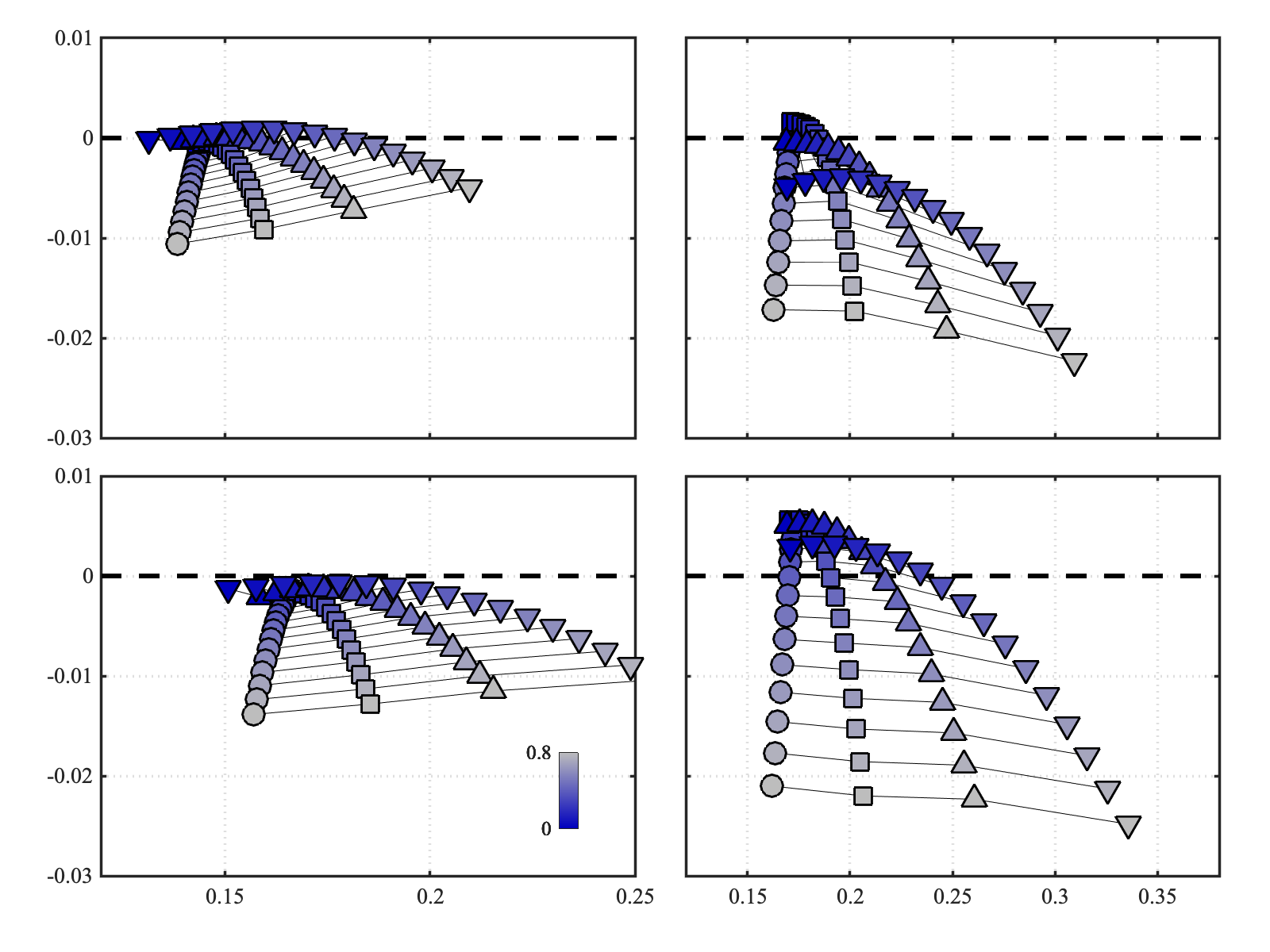}};
		\node[text width=5cm] at (5.8,9.4) {$\alpha = 16^\circ$};
		\node[text width=5cm] at (5.8,4.7) {$\alpha = 20^\circ$};
		\node[text width=5cm] at (12.3,9.4) {$\alpha = 26^\circ$};
		\node[text width=5cm] at (12.3,4.7) {$\alpha = 30^\circ$};
		\node[text width=5cm] at (14.0,8.35) {\color{blue2}\textbf{stable}};
		\node[text width=5cm] at (13.8,9.00) {\color{green2}\textbf{unstable}};
		\node[text width=5cm] at (5.1,6.35) {\Large $\circ$};
		\node[text width=5cm] at (5.4,6.4) {$\Lambda = 0^\circ$};
		\node[text width=5cm] at (5.1,5.97) {\scriptsize $\square$};
		\node[text width=5cm] at (5.4,6.0) {$\Lambda = 15^\circ$};
		\node[text width=5cm] at (6.8,6.35) {\scriptsize $\triangle$};
		\node[text width=5cm] at (7.1,6.4) {$\Lambda = 30^\circ$};
		\node[text width=5cm] at (6.8,5.97) {\normalsize $\triangledown$};
		\node[text width=5cm] at (7.1,6.0) {$\Lambda = 45^\circ$};
		\node[text width=5cm] at (2.5,7.45) {\rotatebox{90}{$St_i^\prime$}};
		\node[text width=5cm] at (2.5,2.8) {\rotatebox{90}{$St_i^\prime$}};
		\node[text width=5cm] at (6.0,0.15) {\rotatebox{0}{$St_r^\prime$}};
		\node[text width=5cm] at (12.4,0.15) {\rotatebox{0}{$St_r^\prime$}};
		\scriptsize
		\node[text width=0cm] at (5.8,2.4) {${k_{z^\prime} (\pi)}$};
		\end{tikzpicture} \vspace{-5mm}
		\caption{Dominant eigenvalues of $\mathbf{L}_{\bar{\mathbf{q}}}$ for (a) $\alpha = 16^\circ$, (b) $20^\circ$, (c) $26^\circ$, and (d) $30^\circ$, for different sweep angles $\Lambda$, and spanwise wavenumbers $k_{z^\prime}$. $St_r^\prime$ and $St_i^\prime$ are the $St^\prime$ numbers for growth rate and temporal frequency, respectively. Black solid lines connect the eigenvalues for the same $k_{z^\prime}$.} 
		\label{fig:eigsPathforAoA}
	\end{figure}
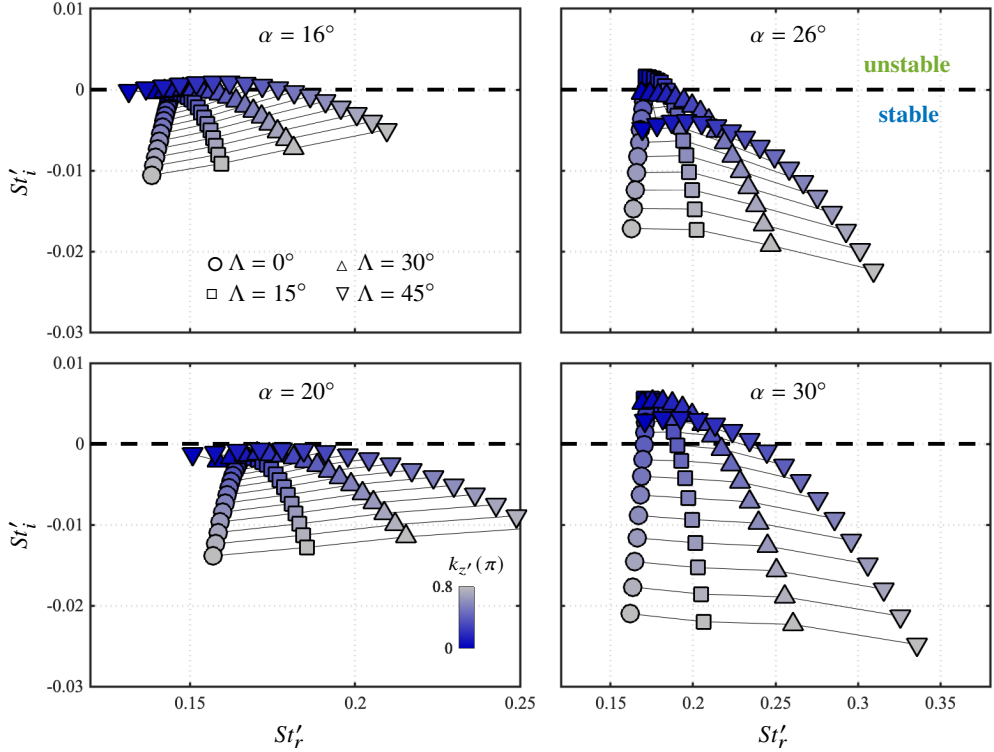

	The eigenspectrum of the linearized Navier--Stokes operator $\mathbf{L}_{\bar{\mathbf{q}}}$ is comprised of eigenvalues $-i \omega = -i \omega_r + \omega_i$, with growth rate $\omega_i$ and temporal frequency $\omega_r$. The dominant eigenmode reveals the spatial structures that can emerge in the flow. We track the dominant eigenvalue in the complex plane as we increase the spanwise wavenumber $k_{z^\prime}$ for each $(\alpha, \Lambda)$ pair to examine if some of these parameters may cause the linear operator to become unstable, as shown in figure \ref{fig:eigsPathforAoA}. 
	
	There is a distinct behavior for $\alpha \le 20^\circ$ and $\alpha \ge 26^\circ$. For the lower angles of attack, swept wings have a greater growth rate for each $k_{z^\prime}$, while for the higher angles of attack we observe the opposite trend. For $\alpha \le 20^\circ$, swept wings wakes are close to the stability threshold as we increase $k_{z^\prime}$. On the other hand, for $\alpha \ge 26^\circ$ unstable modes move into the stable region as we increase $k_{z^\prime}$ for all swept wings. As the linearized operators are unstable for $\alpha = 30^\circ$ and $k_{z^\prime} \approx 0$, small perturbations can be amplified and sustained by the wavemakers generating the $3$D wake flow observed in the numerical simulations. 
	
	Even when the modes are unstable, they are close to the stable region in the complex plane. We keep the same finite time window for resolvent analysis for all angles of sweep and attack. We find the highest growth rate among all cases to set the discounted resolvent operator with a finite time shorter than the associated time scale of the largest $\omega_i$, which is observed for the unswept wing at $\alpha = 30^\circ$. For this reason, we use a fixed time window of $t_sU_\infty \sin\alpha / L_c\cos \Lambda = 50$ for the discounted resolvent analysis of all angles of attack and sweep. 
	
	\section*{Acknowledgments}
	\label{sec:acknowledgments}
	We acknowledge the support from the Air Force Office of Scientific Research (program manager: Dr. G. Abate, grant number: FA9550-21-1-0174). We thank C. S. Skene, T.  R. Ricciardi, M. Amitay, and V. Theofilis for enlightening discussions on wake dynamics and resolvent analysis. Some of the computations herein were supported by the Department of Defense High Performance Computing Modernization Program and the Texas Advanced Computing Center (TACC) at The University of Texas at Austin.  
	
	\section*{Declaration of interest}
	\label{sec:doi}
	The authors report no conflict of interest.
	
	\bibliography{taira_refs}
	\bibliographystyle{jfm}
	
\end{document}